\documentstyle[12pt,axodraw]{article}
\newcommand{\be}{\begin{equation}}
\newcommand{\ee}{\end{equation}}
\newcommand{\bea}{\begin{eqnarray}}
\newcommand{\eea}{\end{eqnarray}}
\newcommand{\non}{\nonumber}

\begin{document}
\thispagestyle{empty}
\onecolumn
\date{March 14, ~~1997}
\vspace{-1.4cm}
\begin{flushleft}
{DESY 96-240 }\\
{JINR E2-96-433 }\\
{hep-ph/9703319}\\

\end{flushleft}
\vspace{1.5cm}

\begin{center}

{\LARGE {\bf

  Generalized recurrence relations for

  two-loop propagator integrals with arbitrary masses}

}

\vspace{1.5cm}

\vfill
{\large
 O.V.~Tarasov \footnote{E-mail:
 $tarasov@ifh.de$ \\
 {~On leave of absence from JINR, 141980 Dubna (Moscow Region), Russian
 Federation.}}}

\vspace{2cm}
Deutsches Electronen-Synchrotron DESY \\
Institut f\"ur Hochenergiephysik IfH, Zeuthen\\
Platanenallee 6, D--15738 Zeuthen, Germany
\vspace{1cm}
\end{center}

\vfill

\begin{abstract}
An  algorithm  for calculating  two-loop propagator type Feynman
diagrams with arbitrary masses and external momentum is proposed.
Recurrence relations allowing to express any scalar integral
in terms of basic integrals are given. A minimal set
consisting of 15 essentially two-loop  and 15 products of 
one-loop  basic integrals is found. Tensor integrals and 
integrals with irreducible numerators are represented as a 
combination of scalar ones with a higher space-time dimension
which are reduced to the basic set  by using the generalized
recurrence relations proposed in \cite{connection}.
\end{abstract}

\vfill
\newpage

\setcounter{footnote}{0}


\section{Introduction}

Mass effects play an important role in confronting
experimental data obtained at the high-energy colliders, like LEP
and SLC, with theoretical predictions.
Precise determinations of many physical parameters in the
Standard  Model (SM) require the evaluation  of the
mass dependent radiative corrections.

At the one-loop level the problem of calculating  Feynman
integrals with several masses was in principle solved in \cite{HV}. 
The history of two-loop calculations is rather long. First results
were reported more than fourty years ago \cite{JL}, \cite{KS}. 
During the last years a number of approaches for the evaluation of
 two-loop diagrams with massive particles were proposed 
 \cite{BoDa}
-- \cite{GV}.
Integrals with specific topologies  or mass combinations 
were considered in \cite{mendels} 
-- \cite{BJ}. Most results for scalar master integrals were obtained
using dispersion relations. The first  generalization of 
the method of integration by parts \cite{Tkachov} to massive
integrals was done for the on-shell
diagrams with one massive particle \cite{GBGS}. The evaluation
of the two-loop correction for the photon propagator with the
aid of recurrence relations was performed in \cite{BFT}.
Up to now all attempts to extend the standard method of 
integration by parts \cite{Tkachov} to two-loop diagrams with
arbitrary external momentum and all  masses different were 
unsuccessful. For a short review of some techniques 
for calculating two-loop diagrams, see for instance Ref. \cite{AD96}.

Despite these efforts  even for  propagator type two-loop
diagrams no satisfactory solution similar to the one-loop
case is available so far.
In the SM, due to the complicated structure of
integrals and the large number of diagrams (typically thousands),
no complete calculation of a two-loop self-energy has been carried
out. Many different species of particles with different masses
have to be taken into account and this makes the evaluation of 
two-loop integrals a rather difficult problem. Existing  numerical
methods for evaluating two-loop diagrams cannot guarantee required
accuracy for the sum of  thousands of  diagrams.
More promising may be the semi-analytic
method proposed in Ref. \cite{FT}. It is
based on momentum expansion, conformal mapping and
the use of Pad\'e approximants.

In calculating Feynman diagrams mainly  three  difficulties
arise: tensor decomposition of integrals,
reduction of  scalar integrals to several basic  integrals
and  the  evaluation of scalar basic integrals.

The first two problems were addressed in Refs. \cite{BW}, \cite{WMSB}
where an algorithm for the analytic calculation of two-loop diagrams
with several masses was proposed.
This algorithm, in  principle, allows one to express any diagram
in terms of several scalar integrals  but in our opinion the problem
was not solved completely.  In calculating the tensor integrals,
the authors of Ref. \cite{BW} have to introduce in   addition to the
original five propagators with different masses 
a fictitious massless propagator. As a consequence, in the final
result there are not only integrals with the original masses but
in general  also integrals having an extra massless propagator.
These integrals may correspond  to diagrams having a topology which
is not inherent to the original  diagram under consideration. 
The minimal set  of independent basic integrals  was not found
in \cite{BW}. In fact, the algorithm of  Ref.\cite{BW}
allows only  partial reduction and reveals  even in the simplest
case with one massive particle an overcomplicated result. 
For example, the two-loop correction to the photon propagator 
in \cite{BW} is given in terms of 4   two-loop 
scalar integrals, though there must be only two \cite{BFT}.
The position of the nearest threshold singularity of the
two additional integrals is not characteristic to the
two-loop photon propagator.
Without further simplification of the result additional integrals
may cause potentially large numerical cancellations.

 Recently in Ref. \cite{GY} a framework for treating all two-loop
diagrams which occur in a renormalizable theory was developed.
To reduce tensor integrals to scalar ones the authors introduced
as a basis ten scalar integrals. For nonrenormalizable theories
they  predict the impossibility to isolate a finite number of
functions from which all other can be obtained by differentiation.
As will be seen from our paper there is no difference between
two-loop propagator integrals occurring in renormalizable and
non-renormalizable theories. Any two-loop, so-called  "London
transport" (or ``sunset'' ) type diagram with arbitrary masses 
considered in \cite{GY} can be expressed as a combination of
four two-loop integrals and products of one-loop tadpole integrals.

 In the present paper we propose a method which completely solves
for propagator diagrams the first of the two  aforementioned problems.
We formulate an algorithm for transforming tensor integrals
into a combination of scalar ones.
In fact, this is a specification to the two-loop case
of the general algorithm given in Ref. \cite{connection} .
  We show how to reduce all scalar integrals
  occurring in the computation of an arbitrary two-loop diagram
  to a sum over a minimal set of basic integrals with rational
  coefficients depending on masses, momenta and the dimensionality
  of space-time  $d$.
All recurrence relations needed for the reduction of a variety
of Feynman integrals to basic integrals are
presented. The minimal set of basic integrals is given.

The paper is organized as follows. In Sec.2 we describe an
optimal procedure for the simplification of the integrand
of a diagram. In Sec. 3 a method for the representation of tensor
integrals and integrals with irreducible numerators in terms of
scalar ones with shifted $d$ is presented.
In Sec. 4 generalized  recurrence relations which are necessary
for the reduction of two-loop integrals to basic ones are given.
In Appendix a complete set of basic integrals is given.

\vspace{2.0cm}

\section{General notation and integrand simplification}

The subject of our consideration will be  two-loop two-point
dimensionally regulated diagrams with arbitrary masses.
In principle, with the method presented in this article
one can treat  the diagram in tensor form. In Sec. 3 an algorithm 
for the tensor decomposition of individual
integrals will be given. Technically the  evaluation of
diagrams in tensor form is more involved than 
the evaluation of scalar diagrams. In practice one need to know
the coefficient of the particular tensor structure of the diagram.
After contracting a diagram with an appropriate projection
operator, performing traces we commonly encounter integrals of 
the form:
\be
I^{(d)}(q^2)=\frac{1}{\pi^d}
\int \!\!\int \frac{d^d k_1 d^dk_2}
 { c_1^{\nu_1} c_2^{\nu_2} c_3^{\nu_3} c_4^{\nu_4} c_5^{\nu_5}}
N(k_1^2,k_2^2,k_1q,k_2q,k_1k_2),
\label{J}
\ee
where $N(k_1^2,k_2^2,k_1q,k_2q,k_1k_2)$ is a polynomial and
\begin{tabbing}
   $c_1=k_1^2-m_1^2+i\epsilon$,\qquad\=
   \qquad\=$c_3=(k_1-q)^2-m_3^2+i\epsilon$,
          \qquad\=   $c_5=(k_1-k_2)^2-m_5^2+i\epsilon$,\\
      $c_2=k_2^2-m_2^2+i\epsilon$, \>
    \> $c_4=(k_2-q)^2-m_4^2+i\epsilon.~~~~$ \>  \\
\end{tabbing}
For brevity, we shall omit the ``causal''  $i\epsilon$'s below.
Individual integrals can be specified by the dimensionality
of the space-time, $d$ and by the  powers
of denominators $\nu_i$, called indices of the lines.
Figure 1 shows a generic topology of the two-loop two-point
diagram with  lines labeled as  indicated. We assume that to 
the $i$th line here, corresponds a factor $1/c_i^{\nu_i}$. 
In the case of integer $\nu$'s any two-loop diagram with
different topology can also be reduced to integrals of the form
(\ref{J}), if we use the partial fraction decomposition for 
denominators having the same momenta but different masses.

\vspace{1.8cm}

\begin{center}
\begin{picture}(440,100)(0,0)
\CArc(190,110)(30,0,180)
\CArc(190,110)(30,180,360)
\Line(190,80)(190,140)
\ArrowLine(140,110)(160,110)
\ArrowLine(220,110)(240,110)
\Vertex(160,110){1.75}
\Vertex(190,80){1.75}
\Vertex(190,140){1.75}
\Vertex(220,110){1.75}
\Text(160,130)[]{1}
\Text(220,130)[]{2}
\Text(160,90)[]{3}
\Text(220,90)[]{4}
\Text(197,110)[]{5}
\Text(146,117)[]{$q$}
\Text(140,50)[]{${\rm~~~~~~~~~~~~~~~~~~~~
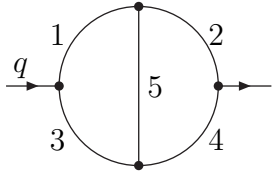
~Fig.~1.~ The~ master~ two-loop~ two-point~ diagram}$}
\end{picture}
\end{center}
\vspace{-1.0cm}
The evaluation of the integrals (\ref{J}) will be performed in
several steps. First, we simplify the integrand as much as possible,
by bringing it to a standard form suitable for further calculations.
Second, one has to get rid off the  numerator
$N(k_1^2,k_2^2,k_1q,k_2q,k_1k_2)$
by introducing  scalar integrals with a higher space-time 
dimension whenever needed. Third, by turns, two-loop integrals with 
five, four and three  lines having different powers of propagators
will be reduced to a set of basic integrals. Fourth, integrals
in  higher dimensions must be reduced to a basic set of 
integrals in the generic dimension $d$. Finally, one-loop
propagator and tadpole integrals should be evaluated.

Putting the integrand into a useful form is just a
matter of tedious algebra. We start the simplification
with the repeated use of the substitutions:
\bea
\frac{(k_1k_2)^{\alpha}}{c_5^{\nu_5}}
&=& \frac{(k_1k_2)^{\alpha}}{c_5^{\nu_5}}\left(
\frac{k_1^2-c_5+k_2^2-m_5^2}{2~k_1k_2}\right)^{<\alpha,\nu_5>}
 \nonumber \\
\frac{(k_2q)^{\alpha}}{c_4^{\nu_4}}
 &=&
  \frac{(k_2q)^{\alpha}}{c_4^{\nu_4}} \left(
  \frac{k_2^2-c_4+q^2-m_4^2}{2~k_2q}\right)^{<\alpha,\nu_4>}
 \nonumber \\
\frac{(k_1q)^{\alpha}}{c_3^{\nu_3}}
&=&
\frac{ (k_1q)^{\alpha}}{c_3^{\nu_3}}
\left(\frac{k_1^2-c_3+q^2-m_3^2}{2~k_1q}\right)
 ^{<\alpha,\nu_3>} \nonumber \\
\frac{(k_2^2)^{\alpha}}{c_2^{\nu_2}}
&=&
\frac{(k_2^2)^{\alpha}}{c_2^{\nu_2}}
\left(\frac{c_2+m_2^2}{k_2^2} \right)^{<\alpha,\nu_2>}
\nonumber \\
\frac{(k_1^2)^{\alpha}}{c_1^{\nu_1}}
&=&\frac{(k_1^2)^{\alpha}}{c_1^{\nu_1}}
\left(\frac{c_1+m_1^2}{k_1^2} \right)^{<\alpha,\nu_1>},
\eea
where we used the notation:
$$
<\!\alpha,\nu_i\!>={\rm min}(\alpha, \nu_i).
$$
After performing these substitutions scalar products in the numerator
remain only in the case when  at least one line  is
eliminated. For integrals of this kind
the simplification of the integrand proceeds as follows.

The scalar product $(k_1k_2)$  remains in the numerator only
if invariant $c_5$ is cancelled, i.e. the integral is in
fact the product of one-loop tensor integrals. In this case
the substitution:
\be
k_1k_2=A(k_1,k_2)+\frac{(k_1q)(k_2q)}{q^2},
\ee
where
\be
A(k_1,k_2)=k_{1\mu}\left(g_{\mu \nu}-\frac{q_{\mu}q_{\nu}}{q^2}\right)
k_{2\nu},
\label{A}
\ee
allows one to transform the integral into a sum
of products of one-loop scalar integrals. Integrals 
with  $A^l(k_1,k_2)$, where $l$ is odd (i.e. with odd powers
of $k_1$ or $k_2$ in the numerator) are zero since the transverse
tensor standing in braces of Eq.($\ref{A}$) will be always 
multiplied by an external momentum $q$. Integrals with even 
powers of $A(k_1,k_2)$ are:
\bea
&&\int\! \int d^dk_1d^dk_2 f_1(k_1,q,m_i)f_2(k_2,q,m_i)
A^{2n}(k_1,k_2)=\frac{\Gamma(n+\frac12)
\Gamma\left(\frac{d-1}{2}\right)}
 {\Gamma(\frac12)\Gamma\left(n+\frac{d-1}{2}\right)}
\nonumber \\
&& \nonumber \\
&& ~~~\times
\int\! \int d^dk_1d^dk_2 f_1(k_1,q,m_i)f_2(k_2,q,m_i) A^{n}(k_1,k_1)
A^{n}(k_2,k_2).
\eea
Once the factor $A(k_1,k_2)$ is removed the integrations 
with respect to  $k_1$ and $k_2$ are completely decoupled.

The integrand can be further simplified by applying
the substitutions:
\bea
&&\frac{(k_1^2)^{\alpha}}{c_3^{\nu_3}}
 =\frac{(k_1^2)^{\alpha}}{c_3^{\nu_3}}
  ~\left(\frac{c_3+2k_1q-q^2+m_3^2}
     {k_1^2}\right)^{<\alpha,\nu_3>},
\\
&&\frac{(k_2^2)^{\alpha}}{c_4^{\nu_4}}
= \frac{(k_2^2)^{\alpha}}{c_4^{\nu_4}}~
\left(\frac{c_4+2k_2q-q^2+m_4^2}{k_2^2}\right)^{<\alpha, \nu_4>}.
\eea
After these substitutions have been made, $k_1^2$, $k_2^2$ will 
remain in the numerator only in bubble-like integrals. Integrals
with irreducible numerators   reveal themselves as  integrals
having scalar products $(k_1q)^{\alpha}$ and/or $(k_2q)^{\beta}$
in the numerator. In principle it is possible to eliminate 
$(k_1q)$, $(k_2q)$ at the expense of $k_1^2$, $k_2^2$. As we shall 
see later, integrals with $(k_1q)$, $(k_2q)$ will produce scalar 
integrals with a smaller shift of $d$ (and therefore  simpler
expressions for further computation) than integrals with $k_1^2$,
$k_2^2$.

If $c_3$ and $c_4$ are absent in the integrand then the scalar
products $(k_1q)$, $(k_2q)$ can be eliminated.
In this case, for an arbitrary scalar function $f(k_1,k_2)$,
\be
\int \!\int d^dk_1 d^dk_2 f(k_1,k_2) (k_1q)^{\nu_1}(k_2q)^{\nu_2}
=v(\nu_1,\nu_2),
\label{v}
\ee
is zero if $\nu_1+\nu_2$ is odd. For  $\nu_1+ \nu_2$ even,
the scalar products $(k_1q)$, $(k_2q)$ can be eliminated
from the integrand applying the recurrence relations
derived in \cite{ai95}:
\be
(d\!+\nu_1+\nu_2\!-2)v(\nu_1,\nu_2)=q^2 \{ (\nu_1-1)k_1^2{\bf 1^-}
\!+\nu_2(k_1k_2){\bf 2^-}\} {\bf 1^-} \circ v(\nu_1,\nu_2),
~~~(\nu_1>1)
\ee
\be
(d+\nu_1+\nu_2-2)v(\nu_1,\nu_2)=q^2 \{ (\nu_2-1)k_2^2{\bf 2^-}
+\nu_1(k_1k_2){\bf 1^-} \} {\bf 2^-} \circ v(\nu_1,\nu_2),
~~~(\nu_2>1),
\ee
where ${\bf 1^{\pm}}v(\nu_1,\nu_2)\equiv v(\nu_1 \pm 1,\nu_2)$
etc., and the $\circ$ sign means that factors $k_1^2, k_2^2,k_1k_2$ in
braces must be considered under the integral sign in $v$.
To simplify  one-loop tadpole-like integrals, rather
frequently, application of the  simple formula
\be
\int \frac{d^dk_1}{(k_1^2-m_1^2)^{\nu_2}}(2k_1q)^{2\nu_1}
=\frac{(2\nu_1)! }{(d/2)_{\nu_1} (\nu_1)!}(q^2)^{\nu_1}
\int \frac{d^dk_1}{(k_1^2-m_1^2)^{\nu_2}}(k_1^2)^{\nu_1},
\ee
turns out to be useful.

Using the formulas presented in this section, a scalar contribution 
(\ref{J}) will be transformed into a sum of integrals with
irreducible numerators $(k_1q)^{\alpha}(k_2q)^{\beta}$ 
having at maximum four lines and a variety of scalar integrals 
with different indices.
Integrals with irreducible numerators $(k_1q)$, $(k_2q)$
will be considered in the next section.
Notice that, the integrals (\ref{v}) can also be regarded as
integrals with irreducible numerators and they can be
treated according to the formalism discussed in the next section.


\section{Tensor integrals and irreducible numerators}

Tensor integrals can be written as a combination of scalar integrals
with shifted space-time dimension multiplied by tensor structures
made from  external momenta and the metric tensor.
This was shown in Ref.\cite{Andrei} for the one-loop case  and
in Ref.\cite{connection} for an arbitrary case.
In the approach of Ref. \cite{connection}
the reduction of tensor integrals to such a  representation
is performed by applying a tensor operator to a scalar integral.
For the two-loop tensor integrals of interest the following relation
holds:
\be
\int \! \int \frac{d^dk_1 d^dk_2}
  {c_1^{\nu_1}c_2^{\nu_2} c_3^{\nu_3}c_4^{\nu_4}c_5^{\nu_5}}
  ~k_{1\mu_1} \ldots k_{1\mu_r}
  k_{2\lambda_1} \ldots k_{2\lambda_s}
 \non \\
 =T_{\mu_1\ldots \lambda_s} (q,\{{\partial}_j\},
{\bf d^+})
\int \!\! \int \frac{d^dk_1 d^dk_2}
{c_1^{\nu_1} c_2^{\nu_2} c_3^{\nu_3} c_4^{\nu_4} c_5^{\nu_5}},
\label{tensint}
\ee
where
$$
\partial_j=\frac{\partial}{\partial m_j^2},
$$
and ${\bf d^+}$ is the operator shifting the value of the space-time
dimension of the integral by two-units:
${\bf d^+} I^{(d)}=I^{(d+2)}$.
On the right-hand side of Eq.(\ref{tensint}) it is assumed
that, at the beginning,  invariants $c_i$ have different nonzero
masses and after differentiation with respect to $m_i^2$ they must 
be set to their original values.

To illustrate the method of Ref. \cite{connection}
we will derive here an explicit expression for the tensor operator
$T_{\mu_1\ldots \lambda_s} (q,\{{\partial}_j\},{\bf d^+}) $.
The main ingredients of the derivation are independent auxiliary
vectors $a_1, a_2$ and the  use of the  $\alpha$- parametric
representation \cite{BS}. The tensor structure of the integrand on
the left-hand side of (\ref{tensint})  can be written as
\be
k_{1\mu_1} \ldots k_{1\mu_r}   k_{2\lambda_1} \ldots  k_{2\lambda_s}=
\left.\frac{1}{i^{r+s}}\frac{\partial}{\partial a_{1\mu_1}}
 \ldots \frac{\partial}{\partial a_{1\mu_r}}
 ~\frac{\partial}{\partial a_{2\lambda_1}}
 \ldots \frac{\partial}{\partial a_{2\lambda_s}}
 \exp \left[i (a_1k_1+a_2k_2)\right]
\right|_{ a_i=0 }.
\label{avectors}
\ee
To convert the integral
\be
G^{(d)}(q^2)=\int \! \int \frac{ d^dk_1 d^dk_2}
{c_1^{\nu_1}c_2^{\nu_2}c_3^{\nu_3}c_4^{\nu_4}c_5^{\nu_5}}
\exp \left[i (a_1k_1+a_2k_2) \right].
\ee
into the $\alpha$-parametric representation
we apply standard methods (see for example, \cite{BS}).
Transforming all propagators into a parametric form
\be
\frac{1}{(k^2-m^2+i\epsilon)^{\nu}}
 = \frac{i^{-\nu}}{ \Gamma(\nu)}\int_0^{\infty}
 d\alpha ~\alpha^{\nu-1} \exp\left[i\alpha(k^2-m^2+i\epsilon)\right],
\ee
and using the $d$- dimensional Gaussian integration  formula
\be
\int d^dk \exp \left[i(A k^2+ 2(pk))\right] =i
 \left( \frac{\pi}{i A} \right)^{\frac{d}{2}}
 \exp \left[ -\frac{ip^2}{A} \right] ,
\ee
we can easily evaluate the integrals over loop momenta.
The final result is:
\be
G^{(d)}(q^2)\!=i^2\!\left ( \frac{\pi}{i} \right)^d
\! \prod^{5}_{j=1} \frac{i^{-\nu_j}}{\Gamma(\nu_j)}
\! \int_0^{\infty} \!\!\!\! \ldots \!\! \int_0^{\infty}
\frac{d \alpha_j \alpha^{\nu_j-1}_j}
     { [ D(\alpha) ]^{\frac{d}{2}}}
  \exp \left[
              i \left(\frac{Q(\alpha,a_1,a_2)}
	                   {D(\alpha)}
             \!-\!\sum_{l=1}^{5}\alpha_l(m_l^2\!-\!i\epsilon)
                \right)
         \right],
\label{repres}
\ee
where
\bea
\label{Dform}
&&D(\alpha)=\alpha_5(\alpha_1+\alpha_2+\alpha_3+\alpha_4)
  +(\alpha_1+\alpha_3)(\alpha_2+\alpha_4),    \\
&& \non \\
&&Q(\alpha,a_1,a_2)=[(\alpha_1+\alpha_2)(\alpha_3+\alpha_4)
\alpha_5 +\alpha_1\alpha_2(\alpha_3+\alpha_4)
         +\alpha_3\alpha_4(\alpha_1+\alpha_2)]q^2 \non \\
&&~~~~~~~~~~~~+(qa_1)Q_1+(qa_2)Q_2+a_1^2Q_{11}
            +a_2^2Q_{22}+(a_1a_2)Q_{12},
\eea
and
\bea
Q_1&=&\alpha_3\alpha_5+\alpha_4\alpha_5+\alpha_2\alpha_3
+\alpha_3\alpha_4, \nonumber \\
Q_2&=&\alpha_4\alpha_5+\alpha_3\alpha_5+\alpha_1\alpha_4
+\alpha_3\alpha_4, \nonumber \\
-4 Q_{11}&=&\alpha_2+\alpha_4+\alpha_5,
\nonumber \\
-4Q_{22}&=&\alpha_1+\alpha_3+\alpha_5,
\nonumber \\
-2Q_{12}&=&\alpha_5.
\label{qbeta}
\eea
It is straightforward to work out from (\ref{tensint}),
(\ref{avectors}) and  the explicit  expression ($\ref{repres}$) that
\bea
&&T_{\mu_1 \ldots \lambda_s}(q,\left\{\partial\right\},{\bf d^+})
 =\frac{1}{i^{r+s}}\!\prod_{j=1}^{r} \frac{\partial}
 { \partial a_{1\mu_j}} \!\! \ldots \prod_{n=1}^{s}
 \frac{\partial}{ \partial a_{2 \lambda_n}}
 \non \\
&&~~~\times \exp
 \left[i \left(
 (qa_1)Q_1+(qa_2)Q_2+a_1^2Q_{11}+a_2^2Q_{22}+(a_1a_2)Q_{12}
 \right) \rho \right]
 \left|_{ a_j=0 \atop {\alpha_j=i \partial_j \atop
  \rho=-\frac{1}{\pi^2}{\bf d^+}} } \right. .
\label{Ttensor}
\eea
This operator is a particular case of the more general one
derived in Ref. \cite{connection}. Formula (\ref{tensint})
can be used for the direct evaluation of two-loop tensor
integrals. Notice that in order to obtain the tensor
decomposition of the integral no contractions with external 
momenta and the metric tensor and no solution of a linear 
system of equations are needed.

Integrals with irreducible numerators
\be
I^{(d)}_{r s}= \int \!\! \int \frac{d^dk_1 d^dk_2}
{c_1^{\nu_1} c_2^{\nu_2} c_3^{\nu_3}c_4^{\nu_4} c_5^{\nu_5}}
(k_1q)^{r}(k_2q)^{s},
\label{irred}
\ee
 can be regarded as a contraction of the tensor integral
 (\ref{tensint}) with $q_{\mu_1}\ldots q_{\nu_s}$.
For the scalar integrals (\ref{irred}) a somewhat simpler
formula, analogous to (\ref{tensint}), can be derived
by introducing auxiliary {\it scalar} parameters $\beta_j$.
Similar to (\ref{avectors}) we write
\be
(k_1q)^{r}(k_2q)^{s}=
\frac{\partial^{r}}{(i\partial \beta_1)^{r}}
\frac{\partial^{s}}{(i\partial \beta_2)^{s}}
\exp\left\{ i[\beta_1(k_1q)+\beta_2(k_2q)] \right\}
\left|_{\beta_i=0}\right. .
\label{Irs}
\ee
Carrying through the steps which were leading us to (\ref{Ttensor}),
we find the relation:
\be
I^{(d)}_{r s}=T_{rs} (q,\{{\partial}_j\},{\bf d^+})
I^{(d)}_{0 0},
\label{form23}
\ee
where
\bea
&&T_{r s}(q,\{{\partial}_j\},{\bf d^+})=
\frac{1}{i^{r+s}}
\frac{\partial^{r}}{\partial \beta_1^{r}}
 \frac{\partial^{s}}{\partial \beta_2^{s}}
\non \\
&&~~~~~~~
 \times \exp\left\{i q^2 [
Q_{1}\beta_1+Q_{2}\beta_2
+Q_{11}\beta_1^2+Q_{22}\beta_2^2+Q_{12}\beta_1 \beta_2
      ]\rho\right\}
 \left|_{{\beta_i=0}
 \atop {\alpha_j=i\partial_j \atop
 \rho=-\frac{1}{\pi^2}  {\bf d^+}} }\right. ,
\label{Tbeta}
\eea
with $Q_i, Q_{ij}$  given in (\ref{qbeta}).

The evaluation of the scalar integrals in the form (\ref{form23})
is more efficient than the use of its tensor analog (\ref{tensint}).
Notice that, as we mentioned in the previous section, irreducible
numerators can appear only if at least one of lines 1 to 4 is 
contracted. In this case the operator
$T_{rs}(q,\{{\partial}_j\},{\bf d^+})$ becomes simpler
because one can set equal to zero  those $\alpha$ parameters  which
correspond to  contracted lines. One can easily build up 
transformation operators 
$T_{\mu_1\ldots \lambda_s} (q,\{{\partial}_j\}$
or $T_{rs}(q,\{{\partial}_j\},{\bf d^+})$ by using
any computer algebra system. Running FORM  \cite{FORM} 
on a PC Pentium 90 it takes usually several seconds (or minutes
in complicated cases) to construct these operators.

As  an application of the formalism presented in  this section,
let us consider a typical integral with an irreducible numerator:
\be
I_{11}=\int \!\! \int \frac{d^dk_1 d^dk_2}{c_2c_3c_5}(k_1q)(k_2q).
\label{example}
\ee
For this particular case,  dropping irrelevant terms with
$\beta_1^2, \beta_2^2$ in (\ref{Tbeta}),
the transformation operator reads:
\be
T_{11}=-\frac{\partial}{\partial \beta_1}
\frac{\partial}{\partial \beta_2}
 \exp \left\{ i q^2 \left[ \beta_1Q_1+\beta_2Q_2+\beta_1\beta_2
 Q_{12} \right]\rho \right\}
 \left|_{\beta_i=0
 \atop {\alpha_j=i\partial_j \atop
 \rho=-\frac{1}{\pi^2} {\bf d^+}} }\right..
\label{Tinexample}
\ee
Since $c_1$ and $c_4$ are absent in the integrand (i.e. 
corresponding lines are contracted), we have to set 
$\alpha_1=\alpha_4=0$  and thus obtain from
(\ref{qbeta}):
\be
Q_1=\alpha_3(\alpha_2+\alpha_5),~~~~~~~~~
Q_2=\alpha_3\alpha_5,~~~~~~~~~
Q_{12}=-\frac12\alpha_5.
\label{Qinexample}
\ee
Substituting these expressions into (\ref{Tinexample}) we get
\be
T_{11}=\frac{q^2}{2 \pi^2}{\bf d^+} \partial_5
+\frac{q^4}{\pi^4}({\bf d^+})^2\partial_3^2\partial_5
(\partial_2+\partial_5).
\ee
With this operator (\ref{form23}) leads to the desired relation:
\bea
&&\int \!\! \int \!\!d^dk_1 d^dk_2
 \frac{(k_1q)(k_2q)}{c_2c_3c_5}\!=
 \nonumber \\
&&\nonumber \\
&&~~~~~~~~~~~~~
\!
\frac{q^2}{2\pi^2} \int \!\! \int \!\!
\frac{d^{d+2}k_1 d^{d+2}k_2}{c_2c_3c_5^2}
+\frac{q^4}{\pi^4}
\int \!\! \int \!\! d^{d+4}k_1 d^{d+4}k_2\left[
\frac{2}{c_2^2c_3^3c_5^2}
+\frac{4}{c_2c_3^3c_5^3}            \right].
\label{examp}
\eea
Integrals on the right-hand side of (\ref{examp}) can be
reduced to basic ones in the generic dimension $d$ by using
recurrence  relations given in the next section.
Notice that for these integrals at arbitrary $d$ an 
analytic expression in terms of Lauricella functions is available
\cite{BBBS} and therefore formula (\ref{Irs}) shows that 
the ``sunset''  type diagrams with any number of scalar products
in the numerator are always expressible in terms of restricted number
of the Lauricella functions.

\section{Recurrence relations}
In the present section we will be concerned with generalized
recurrence relations. Applying methods presented in previous 
sections we have reduced the problem of calculating two-loop
diagrams to that of evaluating scalar integrals without numerator
and having different shifts of $d$. The reduction of scalar 
integrals with different powers of propagators (or  different
indices) to a sum over a minimal set of basic integrals in an 
arbitrary dimension will be done by means of generalized  recurrence
relations.  The method of their derivation was described in 
Ref. \cite{connection}. For the two-loop general mass case the
derivation of these  relations is rather tedious and for brevity
of the presentation will be omitted in the present paper. We will
give only the final formulae and  describe their application.
To avoid overcomplication by irrelevant indices  it is  convenient
to introduce a separate notation for two-loop integrals with 
five, four and three propagators:
\bea
&&  F^{(d)}_{\nu_1 \nu_2 \nu_3 \nu_4 \nu_5}=
\frac{1}{\pi^d}
\int\!\int
\frac{d^d k_1 d^dk_2}{[k_1^2-m_1^2]^{\nu_1}[k_2^2-m_2^2]^{\nu_2}}
\non \\
&&~~~~~~~~~\times \frac{1}
{[(k_1-q)^2-m_3^2]^{\nu_3} [(k_2-q)^2-m_4^2]^{\nu_4}
[(k_1-k_2)^2-m_5^2]^{\nu_5}},
\nonumber \\
&& \non \\
&&  V^{(d)}_{\nu_1 \nu_2 \nu_3 \nu_4}=
\frac{1}{\pi^d}
\int\!\! \int \frac{d^d k_1 d^dk_2}{[(k_1-k_2)^2-m_1^2]^{\nu_1}}
\nonumber \\
&&~~~~~~~~~\times\frac{1}{[k_2^2-m_2^2]^{\nu_2}
[(k_1-q)^2-m_3^2]^{\nu_3} [(k_2-q)^2-m_4^2]^{\nu_4}},
\nonumber \\
&&\non \\
&&  J^{(d)}_{\nu_1 \nu_2 \nu_3}(q^2)=
\frac{1}{\pi^d}
\int\!\! \int \frac{d^d k_1 d^dk_2}
{[k_1^2-m_1^2]^{\nu_1}[(k_1-k_2)^2-m_2^2]^{\nu_2}
[(k_2-q)^2-m_3^2]^{\nu_3}}.
\eea
One-loop integrals will be denoted as
\bea
&&G^{(d)}_{\nu_1 \nu_2 }=\frac{1}{\pi^{\frac{d}{2}} }
\int \frac{d^d k_1}
{[k_1^2-m_1^2]^{\nu_1}[(k_1-q)^2-m_2^2]^{\nu_2}}, \\
&& T^{(d)}_{\nu_1  }= \frac{1}{\pi^{\frac{d}{2}}}
\int \frac{d^d k_1} {[k_1^2-m_1^2]^{\nu_1}}.
\eea
Where no confusion can arise, we simply refer to the
above functions as $F^{(d)}$, $V^{(d)}$, $\ldots$ etc.

The application of recurrence relations to $F^{(d)}$ will produce
$V^{(d)}$, $J^{(d)}$ and more simple one-loop integrals. In turn
integrals $V^{(d)}$ will produce $J^{(d)}$ plus one-loop integrals.
Thus one should first apply the recurrence relations  to $F^{(d)}$,
then to $V^{(d)}$  and after that to $J^{(d)}$.

\subsection{Integrals with five propagators $F^{(d)}$ }

The integrals $F^{(d)}$ originate from  diagrams with the topology
given in Fig. 1. The first step will be  the reduction of indices
$\nu_1, \ldots ,\nu_4$ by using the recurrence relations which
derive from the generalization of the method of integration by parts
\cite{Tkachov} to massive integrals. It is particularly convenient
to represent the recurrence relations in terms of shift operators
decreasing or increasing indices of the integral by one unit
\cite{Tkachov}:
\be
{\bf 1^{\pm}}
 F^{(d)}_{\nu_1\nu_2\nu_3\nu_4\nu_5}
=F^{(d)}_{\nu_1\pm1~\nu_2\nu_3\nu_4\nu_5},~~~~~{\rm etc.}
\ee
with a similar convention for $V^{(d)}$, $J^{(d)}$ and other 
integrals. Notice that an analogous notation we already employed
in Section 2.

The indices  $\nu_1,\ldots ,\nu_4$ can be reduced to 0 or 1 by
iteratively applying the relation:
\bea
&&2 \nu_1 \Delta {\bf 1^+}F^{(d)}_{\nu_1\nu_2\nu_3\nu_4\nu_5}=
 \left\{ (d-2\nu_1-\nu_3-\nu_5)\Delta_1
    +\Delta_{345}
    [\nu_5{\bf 5^+}({\bf 2^-}-{\bf 1^-})-\nu_3{\bf 3^+}{\bf 1^-}]
 \right. \nonumber \\
&& \nonumber \\
&&~~~~~~+\Delta_2[\nu_1{\bf 1^+}({\bf 5^-}-{\bf 2^-})
 +\nu_3{\bf 3^+}({\bf 5^-}-{\bf 4^-})+\nu_5-\nu_1]
\nonumber \\
&& \nonumber \\
&&~~~~~~\left.
     +\Delta_6[\nu_1{\bf 1^+}{\bf 3^-}
     +\nu_5{\bf 5^+}({\bf 3^-}-{\bf 4^-})+\nu_3-\nu_1]
  \right\}F^{(d)}_{\nu_1\nu_2\nu_3\nu_4\nu_5},
\label{line1}
\eea
along with  relations for ${\bf 2^{\pm}}F^{(d)}$,
${\bf 3^{\pm}}F^{(d)}$, ${\bf 4^{\pm}}F^{(d)}$ 
following  from (\ref{line1}) by replacements:
\bea
&&{\bf 2^+}F:~(\nu_1,\nu_3,m_1,m_3,{\bf 1^\pm},{\bf 3^\pm}) \leftrightarrow
              (\nu_2,\nu_4,m_2,m_4,{\bf 2^\pm},{\bf 4^\pm}), \nonumber \\
&&{\bf 3^+}F:~(\nu_1,\nu_2,m_1,m_2,{\bf 1^\pm},{\bf 2^\pm}) \leftrightarrow
              (\nu_3,\nu_4,m_3,m_4,{\bf 3^\pm},{\bf 4^\pm}), \nonumber \\
&&{\bf 4^+}F:~(\nu_1,\nu_2,m_1,m_2,{\bf 1^\pm},{\bf 2^\pm}) \leftrightarrow
              (\nu_4,\nu_3,m_4,m_3,{\bf 4^\pm},{\bf 3^\pm}),
\label{exchanges}
\eea
which are due to the symmetry of the integral.
In the above formula we used the following abbreviations:
\bea
\label{Cayley}
&&\Delta=q^4m_5^2-[(m_1^2+m_2^2+m_3^2+m_4^2-m_5^2)m_5^2
-(m_4^2-m_3^2)(m_2^2-m_1^2)]q^2 \nonumber \\
&&\nonumber \\
&&~~~+(m_3^2-m_1^2)(m_4^2-m_2^2)m_5^2
       +(m_1^2-m_2^2-m_3^2+m_4^2)(m_1^2m_4^2-m_2^2m_3^2),
\eea
\bea
&&\Delta_i= \frac{\partial \Delta}{\partial m_i^2},
\nonumber \\
&& \nonumber \\
&&\Delta_6= \frac{\partial \Delta}{\partial q^2}
=2q^2m_5^2-m_5^2(m_1^2+m_2^2+m_3^2+m_4^2-m_5^2)
+(m_4^2-m_3^2)(m_2^2-m_1^2),
\nonumber \\
&& \nonumber \\
&&\Delta_{ijk}=m_i^4+m_j^4+m_k^4-2(m_i^2m_j^2+m_i^2m_k^2+m_j^2m_k^2)
\non \\
&&~~~~~~=(m_i+m_j+m_k)(m_i+m_j-m_k)(m_i-m_j+m_k)(m_i-m_j-m_k) \non \\
&&~~~~~~=-u_{ijk}(u_{jik}+u_{kij})-u_{jik}u_{kij},\\
&&u_{ijk}=\frac12 \partial_i \Delta_{ijk}=m_i^2-m_j^2-m_k^2,
\non \\
&& \nonumber \\
&&\Delta_{ij6}=m_i^4+m_j^4+q^4-2(m_i^2m_j^2+m_i^2q^2+m_j^2q^2).
\eea
It is worthwhile to note that the repeatedly appearing expression
(\ref{Cayley}) is the well known Cayley kinematical determinant:
\be
\Delta=- \frac12
\left|
\begin{array}{ccccc}
0&1&1&1&1\\
1&0&q^2&m_4^2&m_3^2\\
1&q^2&0&m_2^2&m_1^2\\
1&m_4^2&m_2^2&0&m_5^2\\
1&m_3^2&m_1^2&m_5^2&0
\end{array}
\right|.
\label{determinant}
\ee
To reduce the index $\nu_5$ another recurrence relation is 
needed. It has a form similar to (\ref{line1})
\bea
&&2 \nu_5 \Delta {\bf 5^+} F^{(d)}_{\nu_1\nu_2\nu_3\nu_4\nu_5}=
 \left\{
    \Delta_{136}
    [\nu_3{\bf 3^+}({\bf 4^-}-{\bf 5^-})+\nu_1{\bf 1^+}
    ({\bf 2^-}-{\bf 5^-})]
 \right. \nonumber \\
&& \nonumber \\
&&~~~~~~+(d-\nu_1-\nu_3-2\nu_5)\Delta_5
 +\Delta_2[\nu_5{\bf 5^+}({\bf 1^-}-{\bf 2^-})
 +\nu_3{\bf 3^+}{\bf 1^-}+\nu_1-\nu_5]
\nonumber \\
&& \nonumber \\
&&~~~~~~\left.
     +\Delta_4[\nu_1{\bf 1^+}{\bf 3^-}
     +\nu_5{\bf 5^+}({\bf 3^-}-{\bf 4^-})+\nu_3-\nu_5]
  \right\}F^{(d)}_{\nu_1\nu_2\nu_3\nu_4\nu_5}.
\label{line5}
\eea

If for some mass values $\Delta=0$, then the formulas ($\ref{line1}$),
($\ref{line5}$) are of no use.  This case needs a separate treatment.
For arbitrary $q^2$, the relation $\Delta=0$ can be fulfilled
only at
\be
m_5^2=0,~~~m_4^2=m_3^2,~~~m_2^2=m_1^2.
\label{delta0}
\ee
If this condition is satisfied, substituting (\ref{delta0}) into
($\ref{line1}$) (or ($\ref{line5}$)) yields the simpler relations:
\bea
\label{line5a}
(d-\nu_1-\nu_3-2\nu_5)F^{(d)}_{\nu_1\nu_2\nu_3\nu_4\nu_5}=
[\nu_1{\bf 1^+}({\bf 5^-}-{\bf 2^-})
        +\nu_3{\bf 3^+}({\bf 5^-}-{\bf 4^-})]
F^{(d)}_{\nu_1\nu_2\nu_3\nu_4\nu_5}, \\
 \non \\
(d-\nu_2-\nu_4-2\nu_5)F^{(d)}_{\nu_1\nu_2\nu_3\nu_4\nu_5}=
[\nu_2{\bf 2^+}({\bf 5^-}-{\bf 1^-})
        +\nu_4{\bf 4^+}({\bf 5^-}-{\bf 3^-})]
F^{(d)}_{\nu_1\nu_2\nu_3\nu_4\nu_5}.
\label{line5b}
\eea
By repeated use of these relations, the integrals $F^{(d)}$ can be
reduced to a combination of integrals $J^{(d)}$ and one-loop ones.
Prior to applying (\ref{line5a}), (\ref{line5b}), in some cases, it
may be more efficient to first use the following relation which 
reduces the sum of indices of the integral
\bea
&&
\Delta_{136} {\bf 1^+}F^{(d)}_{\nu_1\nu_2\nu_3\nu_4\nu_5}=
\left\{ u_{613} [\nu_1 {\bf 1^+}{\bf 3^-}
     +\nu_5 {\bf 5^+}({\bf 3^-}-{\bf 4^-})]
 \right. \nonumber \\
&& \nonumber \\
&&~~~~-2u_{136} [\nu_2 {\bf 2^+}({\bf 5^-}-{\bf 1^-})
+\nu_4 {\bf 4^+}({\bf 5^-}-{\bf 3^-})-\Sigma_5-\nu_1+3\nu_3]
\nonumber \\
&& \nonumber \\
&&~~~~
\left.+2m_3^2[\nu_5 {\bf 5^+}({\bf 1^-}-{\bf 2^-})
   +\nu_3 {\bf 3^+}{\bf 1^-}+\nu_1-\nu_3]
\right\}F^{(d)}_{\nu_1\nu_2\nu_3\nu_4\nu_5}.
\label{line1a}
\eea
Analogous recurrence relations for  ${\bf 2^+}F^{(d)}$,
${\bf 3^+}F^{(d)}$, ${\bf 4^+}F^{(d)}$ follow from
(\ref{line1a}) and substitutions  (\ref{exchanges}).
Here, and in the  sequel, we use the notation
\be
\Sigma_i=3d-2\sum_{j=0}^i\nu_j-2.
\ee
Also for particular values of the external momentum  squared $F^{(d)}$
can be reduced to simpler integrals. Again this happens if the
determinant (\ref{determinant}) vanishes.
If $m_5 \neq 0$, then $\Delta=0$ is a quadratic equation for $q^2$
with solutions
\be
q^2_{\pm}=m_1^2+m_3^2-\frac{1}{2m_5^2}[u_{215}u_{435}
 \pm \sqrt{\Delta_{215} \Delta_{435}} ~].
\ee
Furthermore, if $m_5^2=0, m_1 \neq m_2, m_3 \neq m_4$ then we
have $\Delta=0$ at
\be
q^2=\frac{(m_1^2-m_2^2-m_3^2+m_4^2)(m_1^2m_4^2-m_2^2m_3^2)}
         {(m_2^2-m_1^2)(m_3^2-m_4^2)}.
\ee
At $\Delta=0$ the left-hand sides of Eqs. (\ref{line1}),
(\ref{line5}) are zero and we are left with  recurrence
relations of a form similar to (\ref{line5a}), (\ref{line5b}).
By repeated application of this kind of relations $F^{(d)}$ 
again is reducible to $J^{(d)}$ and one-loop integrals.
For example, the special point $m_1=m_4=0, m_2=m_3=m_5=m$ is the
on-shell value of the momentum
$$
q^2=m^2.
$$
This kind of integrals are encountered in calculation of the
on-shell diagrams for the fermion propagator of QED \cite{GBGS}.
Simplification of integrals at some specific values of $q^2$
can be useful, for example, in establishing gauge invariance
of physical amplitudes.
Considering special cases we have been tacitly assuming
that the dimensional regularization does not break down
at the considered specific values of $q^2$.

Now we proceed with a few remarks concerning a reduction of the 
space-time dimension $d$. If we simplify the numerator of a scalar
integral according to the rules described in Sec. 2 then the
integrals $F^{(d)}$ will occur in the final result without change of 
dimension. Decomposing   tensor integrals  we will encounter  
in  general $F^{(d)}$'s with shifted $d$.   The connection between
$F^{(d)}_{11111}$ and $F^{(d-2)}_{11111}$ which suffices to
reduce these integrals to generic dimension can be worked out 
from the  relation derived in Ref.  \cite{connection}:
\be
F^{(d-2)}_{\nu_1 \nu_2 \nu_3 \nu_4 \nu_5}=
D(\partial)F^{(d)}_{\nu_1 \nu_2 \nu_3 \nu_4 \nu_5}.
\label{changedinF}
\ee
Here $D(\partial)$ is a differential operator obtained from
(\ref{Dform}) by converting $\alpha_j$ into $\partial_j$.
By applying the recurrence relations already presented in this section
Eq. (\ref{changedinF}) can be written in the following form:
\be
\Delta F^{(d-2)}_{11111}=-q^2(d-3)(d-4)F^{(d)}_{11111}
+(P_1 {\bf 1^-}+P_2 {\bf 2^-}+P_3 {\bf 3^-}+P_4 {\bf 4^-}+
P_5 {\bf 5^-})F^{(d)}_{11111},
\label{fddm2}
\ee
where $P_i$ are polynomials in masses, $q^2$, $d$ and operators
${\bf 1^+}$,..,${\bf 5^+}$. The polynomial $P_i$ does not depend on
the operator ${\bf i^+}$. The final formula connecting
$F^{(d)}_{11111}$ and $F^{(d-2)}_{11111}$ with all masses arbitrary
is rather long and for this reason it will be not given here. The
reader can easily obtain it by using the recurrence relations given in
the present article. Notice that at $\Delta=0$ formula (\ref{fddm2})
gives us a relation for $F^{(d)}_{11111}$ in terms of simpler
integrals.

To conclude this subsection, we note that in the case of arbitrary
masses and external momenta by applying ($\ref{line1}$), 
($\ref{line5}$), integrals  $F^{(d)}_{\nu_1\nu_2\nu_3\nu_4\nu_5}$
with integer $\nu_j>0$ will be   expressed   in terms of
$F^{(d)}_{11111}$, $V^{(d)}_{\nu_1\nu_2\nu_3\nu_4}$,
$J^{(d)}_{\nu_1\nu_2\nu_3}$ and more simple ones. Notice that only
one integral with five propagators, namely $F^{(d)}_{11111}$ is 
irreducible with respect to more simple ones. It will be the first
representative of our minimal basis of two-loop integrals.
As mentioned previously in the special case of (\ref{delta0})  all
integrals $F^{(d)}$ and $V^{(d)}$ are reducible to  $J^{(d)}$ and
one-loop ones by means of (\ref{line5a}).

\subsection{Integrals with four propagators $V^{(d)}$}

Now we shall consider the integrals $V^{(d)}$ which at first sight
look simpler than the integrals $F^{(d)}$, but in fact their 
reduction causes more problems. The reason is twofold. First, the
$V^{(d)}$'s are less symmetrical than the $F^{(d)}$'s. Therefore the
number of different relations needed is larger. Second, there are
more cases to be distinguished because  different mass combinations
lead to different recurrence relations. The topology of the diagram
corresponding to the integrals $V^{(d)}$ and the labeling of lines
are shown in Fig. 2.


\begin{center}
\begin{picture}(140,70)(0,0)
\CArc(70,40)(25,0,180)
\CArc(70,40)(25,180,360)
\CArc(45,15)(25,0,90)
\Line(30,40)(45,40)
\Line(95,40)(110,40)
\Vertex(95,40){2}
\Vertex(45,40){2}
\Vertex(70,15){2}
\Text(42,57)[]{$2$}
\Text(42,27)[]{$3$}
\Text(97,27)[]{$4$}
\Text(68,38)[]{$1$}
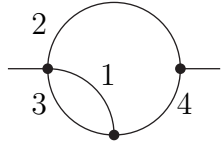
\Text(70,0)[]{${\rm Fig.~2.~Two-loop~diagram~with ~four~propagators}$}
\end{picture}
\end{center}
In many cases the most efficient way to reduce integrals is to first
apply the recurrence relations which  reduce the sum of indices of
the integral. Relations of this kind, which reduce $\nu_{1,2}$ to 0
or 1, are
\bea
&&\Delta_{134}\nu_1 {\bf 1^+}V^{(d)}_{\nu_1\nu_2\nu_3\nu_4}=
\left\{
  2m_3^2\nu_3{\bf 3^+}({\bf 1^-}-{\bf 4^-})+u_{413}\nu_1
  {\bf 1^+}({\bf 3^-}-{\bf 4^-})
\right. \nonumber \\
&& \nonumber \\
&&~~~~~~~~
\left.
 +u_{134}(d-\nu_1-2\nu_3)+2m_3^2(\nu_1-\nu_3)
\right\} V^{(d)}_{\nu_1\nu_2\nu_3\nu_4},
\label{line1v4}
\eea

\bea
&&\Delta_{246}\nu_2 {\bf 2^+}V^{(d)}_{\nu_1\nu_2\nu_3\nu_4}=
 \left\{ 2m_1^2\nu_1{\bf 1^+}({\bf 4^-}-{\bf 2^-})
   +2m_3^2\nu_3{\bf 3^+}({\bf 4^-}-{\bf 2^-})
 \right. \nonumber \\
&& \nonumber \\
&&~~~   -2m_4^2\nu_4{\bf 2^- 4^+}
+(2m_2^2-u_{426})\nu_2{\bf2^+}{\bf 4^-}-(\Sigma_4+2)
({\bf 4^-}-{\bf 2^-})
\nonumber \\
&& \nonumber \\
&&~~~\left.
             +u_{246}(d-3\nu_2)+2m_4^2(\nu_4-\nu_2)
         \right\} V^{(d)}_{\nu_1\nu_2\nu_3\nu_4}.
\label{line2v4}
\eea
The index $\nu_3$ is lowered to zero or one by using the relation
which follows from ($\ref{line1v4}$) by  interchanging
\be
\nu_1 \leftrightarrow \nu_3,
~~m_1 \leftrightarrow m_3,
~~{\bf 1^{\pm}} \leftrightarrow {\bf 3^{\pm}}.
\label{interch13}
\ee
An alternative way of reducing $\nu_3$ is to use first the relation
\be
2m_3^2\nu_3{\bf 3^+}V^{(d)}_{\nu_1\nu_2\nu_3\nu_4}=
[u_{413}\nu_1{\bf 1^+}
+\nu_1{\bf 1^+}({\bf 4^-}-{\bf 3^-})+d-\nu_1-2\nu_3]
V^{(d)}_{\nu_1\nu_2\nu_3\nu_4},
\label{line3alter}
\ee
and then to apply (\ref{line1v4}).

The reduction of $\nu_4$ is more problematic. It can be done at the
expense of increasing  $\nu_1$ and $\nu_2$ by  applying the relation
\bea
&&2 m_4^2 \nu_4 {\bf 4^+}V^{(d)}_{\nu_1\nu_2\nu_3\nu_4}=
 \left\{u_{314}  \nu_1 {\bf 1^+}+u_{624}\nu_2  {\bf 2^+} \right.
 \nonumber \\
&& \nonumber \\
&&~~~~~~~~
 \left.+
    \nu_1{\bf 1^+}({\bf 3^-}-{\bf 4^-})
    -\nu_2{\bf 2^+}{\bf 4^-}
    +d-\nu_1-\nu_2-2\nu_4
 \right\} V^{(d)}_{\nu_1\nu_2\nu_3\nu_4}.
\label{v4line4}
\eea
In turn, $\nu_{1,2}$ are to be lowered by means of (\ref{line1v4}),
(\ref{line2v4}). As an alternative we may use the more symmetrical
relation:
\bea
&&2 m_4^2 \nu_4 {\bf 4^+}V^{(d)}_{\nu_1\nu_2\nu_3\nu_4}=
 \left\{ -2m_1^2 \nu_1 {\bf 1^+}
   +u_{624}\nu_2  {\bf 2^+}
 \right.
 \nonumber \\
&& \nonumber \\
&&~~~~~~~~
 \left.-2\nu_3m_3^2{\bf 3^+}
    -\nu_2{\bf 2^+}{\bf 4^-}
    +2d-2\nu_1-\nu_2-2\nu_3-2\nu_4
 \right\} V^{(d)}_{\nu_1\nu_2\nu_3\nu_4},
\label{v4line4c}
\eea
which  increases not only $\nu_{1,2}$ but also $\nu_3$.

For arbitrary $q^2$ there are two special cases for which some of
the above relations do not work. Relation (\ref{line1v4}) which
reduces $\nu_{1}$ and the analogous one reducing $\nu_{3}$
are nonapplicable if the masses $m_1,m_3,m_4$ are subject to the
condition
\be
\Delta_{134}=(m_1+m_3+m_4)(m_1+m_3-m_4)
               (m_1-m_3+m_4)(m_1-m_3-m_4)=0.
\label{landau}
\ee
Similarly the relation reducing the index $\nu_4$ cannot be used if
\be
m_4=0.
\ee
These cases require additional investigations.

When all masses are different from zero but satisfy
$\Delta_{134}=0$ we found several recurrence relations which allow
us to reduce all $V^{(d)}$- integrals to simpler ones with at least
one line contracted. For example, one possible relation of this kind
\bea
&&[2m_1^2u_{134}(d-\nu_1-2\nu_3)+(\nu_3-1)u_{413}^2
+4m_1^2m_3^2(\nu_1-\nu_3)]V^{(d)}_{\nu_1\nu_2\nu_3\nu_4}= \non \\
&&\non \\
&&~~~[u_{413}\nu_3{\bf 3^+}{\bf 4^-}({\bf 4^-}-{\bf 1^-})
-4m_1^2m_3^2\nu_3 {\bf 3^+}{\bf 1^-}
+(4m_1^2m_3^2+u^2_{413})\nu_3 {\bf 3^+}{\bf 4^-}
\non \\
&&\non \\
&&~~~+u_{413}(\nu_3-1){\bf 1^-}
 -(d-2\nu_1-\nu_3+1)u_{413}{\bf 3^-} \non \\
&&\non \\
&&~~~~~ 
+(d-2\nu_1-2\nu_3+1)u_{413}{\bf 4^-}]
V^{(d)}_{\nu_1\nu_2\nu_3\nu_4}.
\label{recD0}
\eea
follows from Eq. (\ref{line1v4}) with $\Delta_{134}=0$ and using 
relation
\be
2m_1^2\nu_1{\bf 1^+}V^{(d)}_{\nu_1\nu_2\nu_3\nu_4}=
[u_{413}\nu_3{\bf 3^+}
+\nu_3{\bf 3^+}({\bf 4^-}-{\bf 1^-})+d-\nu_3-2\nu_1]
V^{(d)}_{\nu_1\nu_2\nu_3\nu_4},
\label{line1alter}
\ee
which is obtained from (\ref{line3alter}) performing substitutions
(\ref{interch13}).

Before applying  relation (\ref{recD0}) the indices $\nu_{1,2,4}$
should be lowered to 1 with the help of the relations (\ref{line2v4}),
(\ref{v4line4}) and (\ref{line1alter}).  Then applying the relation
(\ref{recD0}) as often as needed, one line in $V^{(d)}$ will be 
eliminated. Notice that, the term ${\bf 3^+}{\bf 4^-}{\bf 4^-}$ 
in (\ref{recD0}) produces factors in the numerator which are supposed
to be treated by the method proposed in  Section 3.

If one of $m_1^2, m_3^2, m_4^2$ is zero then $\Delta_{134}=0$
requires two other masses to be equal. There are two distinct cases:

{\it Case 1:}~~ at $m_1^2=0$ $(m_3^2=0)$ and $m_3^2=m_4^2$ $(m_1^2=m_4^2)$
any integral with four lines can be reduced to a combination
of simpler integrals with three lines by using the relation:
\be
(d-2\nu_1-\nu_3)V^{(d)}_{\nu_1\nu_2\nu_3\nu_4}=
\nu_3{\bf 3^+}({\bf 1^-}-{\bf 4^-}) V^{(d)}_{\nu_1\nu_2\nu_3\nu_4}.
\ee

{\it Case 2:}~~ if $m_4^2=0$ and $m_1^2 = m_3^2$ then $\Delta_{134}=0$
and the index $\nu_2$ should be lowered to one by means
of (\ref{line2v4}). Then the relation
\bea
&&(q^2-m_2^2)(\nu_1-2\nu_2+2\nu_4+2){\bf 4^+}
V_{\nu_1 \nu_2\nu_3 \nu_4}^{(d)}
=\left\{ \nu_1(m_1^2-u_{612}){\bf 1^+}
+4m_2^2 \nu_2 {\bf 2^+}
\right.
\nonumber \\
&& ~~~~~~+2m_1^2 \nu_3 {\bf 3^+}
+[ \Sigma_4-2m_1^2 \nu_1 {\bf 1^+}
-2m_1^2 \nu_3 {\bf 3^+} ]{\bf 2^-} {\bf 4^+}
\nonumber \\
&&~~~~~~~\left. +\nu_1 (q^2-m_2^2){\bf 1^+}{\bf 4^+}{\bf 3^-}
-\Sigma_4 \right\}
V_{\nu_1 \nu_2\nu_3 \nu_4}^{(d)},
\label{eq55}
\eea
is applied until one of the indices $\nu_{2,3,4}$ will be zero.
After each iteration it is reasonable to reduce $\nu_2$ again,
in order to avoid a zero on the left-hand side of (\ref{eq55}) .

Let us consider now the case $m_4^2=0$ and $\Delta_{134} \neq 0$.
Since $\Delta_{134}\neq 0$ the relation (\ref{line1v4}) and its
analog for ${\bf 3^+} V^{(d)}$ may be used to reduce $\nu_{1,3}$
to 1. Then, instead of ($\ref{v4line4}$), the relation
\bea
&&(q^2-m_2^2)[(m_3^2-m_1^2)(d-2\nu_2-2\nu_3+2\nu_4)
+2m_3^2(\nu_3-\nu_1)] V_{\nu_1 \nu_2\nu_3 \nu_4}^{(d)} =\non \\
&& \non \\
&&~~~ \left\{
(m_1^2-m_3^2)[(2m_1^2\nu_1{\bf 1^+}+2m_3^2\nu_3{\bf 3^+}-\Sigma_4-2)
({\bf 2^-}-{\bf 4^-})-4m_2^2\nu_2{\bf 2^+}{\bf 4^-} ]
      \right. \non \\
&& \non \\
&&~~~\left.
+2(q^2-m_2^2)[m_3^2\nu_3 {\bf 3^+}({\bf 1^-}-{\bf 4^-})
             +m_1^2\nu_1 {\bf 1^+}({\bf 4^-}-{\bf 3^-})]
     \right\}V_{\nu_1 \nu_2\nu_3 \nu_4}^{(d)},
\label{line4zer}
\eea
must be applied until one of the $\nu_i$ will become zero. After each
application of (\ref{line4zer}), it is reasonable to reduce 
$\nu_{1,3}$ to 1, if they have been increased after the use of
(\ref{line4zer}).

Again the integrals $V^{(d)}$ at some special values of $q^2$ admit
a reduction to simpler integrals. These special values are determined
by the equation
\be
\Delta_{246}=0,
\label{del246}
\ee
which has the simple solution
\be
q^2_{\pm}=(m_2\pm m_4)^2.
\ee
At $\Delta_{246}=0$ the left-hand side of (\ref{line2v4}) is zero.
With the help of equation (\ref{v4line4}) we exclude the term
${\bf 2^- 4^+}$ from the right-hand side of (\ref{line2v4})  and
obtain a relation which can be used to eliminate one line in
$V^{(d)}$. 
After each application of this relation one should reduce
indices $\nu_{1,3,4}$ to 1 if they have been increased
by the  previous iteration.

So far, we have  considered only relations connecting integrals
with the same space-time dimension. The whole concept  of the
method presented in this article would not work if we would
not able to express integrals with shifted dimension in terms of 
those in a generic dimension. In fact there are several possibilities
to reduce the dimensionality of an integral. As we shall see, one 
possibility always exists and it amounts to first express the
integrals $V^{(d+2k)}_{\nu_1\nu_2\nu_3\nu_4}$ in terms of
$V^{(d+2k)}_{1111}$ and simpler ones and then to express all
$V^{(d+2k)}_{1111}$ in terms of $V^{(d)}_{1111}$ plus simpler ones.

The other possibility is to reduce simultaneously  the indices and the
space-time dimension. We found several  recurrence relations
of this kind:
\bea
&&4q^2\nu_1\nu_2\nu_3{\bf 1^+}{\bf 2^+}{\bf 3^+}
V_{\nu_1 \nu_2\nu_3 \nu_4}^{(d+2)}= \non \\
&&~~~~~-\left\{ 6d-6\nu_1-5\nu_2-6\nu_3-4\nu_4
-6m_1^2\nu_1{\bf 1^+}
+(u_{624}-4m_2^2)\nu_2{\bf 2^+}
        \right. \non \\
&&~~~~\left.-6m_3\nu_3{\bf 3^+}+2(u_{624}-m_4^2)\nu_4{\bf 4^+}
-\nu_2{\bf 2^+}{\bf 4^-}-2\nu_4{\bf 4^+}{\bf 2^-}
      \right\}V_{\nu_1 \nu_2\nu_3 \nu_4}^{(d)}.
\eea

\bea
\label{eq58}
&&4q^2\nu_1(\nu_1+1)\nu_2{\bf 1^+}{\bf 1^+}{\bf 2^+}
V_{\nu_1 \nu_2\nu_3 \nu_4}^{(d+2)}=\non \\
&&~~~~-\left\{6d-6\nu_1-5\nu_2-6\nu_3-4\nu_4
-2(u_{246}+3m_1^2)\nu_1{\bf 1^+}
      \right.
\non \\
&&~~~~+(u_{624}-4m_2^2)\nu_2{\bf 2^+}-6m_3^2\nu_3{\bf 3^+}
+2(u_{624}-m_4^2)\nu_4{\bf 4^+}
\non \\
&&~~~~\left. +2\nu_1{\bf 1^+}({\bf 4^-}-{\bf 2^-})
-\nu_2{\bf 2^+}{\bf 4^-}-2\nu_4{\bf 4^+}{\bf 2^-}
\right\}V_{\nu_1 \nu_2\nu_3 \nu_4}^{(d)}, \\
&& \non \\
\label{eq59}
&&4q^2m_4^2\nu_1\nu_2(\nu_2+1){\bf 1^+}{\bf 2^+}{\bf 2^+}
V_{\nu_1 \nu_2\nu_3 \nu_4}^{(d+2)}= \non \\
&&~~~~-\left\{ [(m_1^2-m_3^2)(2d-2\nu_1-\nu_2-2\nu_3-2\nu_4)
+m_4^2(2\nu_1+\nu_2-2\nu_3)]
      \right.
\non \\
&&~~~~-u_{134}[2m_1^2\nu_1{\bf 1^+}
 +u_{246}\nu_2{\bf 2^+}]+u_{314}[2m_3^2\nu_3{\bf 3^+}
 +\nu_2{\bf 2^+}{\bf 4^-}]\non \\
&&~~~~\left.
 -2m_4^2(\nu_2{\bf 2^+}+\nu_4{\bf 4^+})({\bf 3^-}-{\bf 1^-})
\right\} V_{\nu_1 \nu_2\nu_3 \nu_4}^{(d)} .
\eea
Exploiting the symmetry of $V^{(d)}$, two additional relations follow
from (\ref{eq58}), (\ref{eq59}) and the interchangements 
(\ref{interch13}). Whenever possible, the above mixed recurrence
relations should be used first. Their application is  more efficient
than the step by step reduction of all $V^{(d)}$'s to master
integrals in different dimensions and then the reduction of master
integrals to the generic dimension.

The key relationship between $V_{1111}^{(d)}$ and $V_{1111}^{(d-2)}$,
which is necessary for the $d-$ recurrences, follows from the 
equation (see Ref. \cite{connection}):
\be
V_{1111}^{(d-2)}=\left(\partial_1 \partial_2+\partial_1 \partial_3+
  \partial_1 \partial_4+\partial_2 \partial_3+
\partial_3 \partial_4 \right)
V_{1111}^{(d)}.
\label{reducedinV}
\ee
To simplify the right-hand side of Eq. (\ref{reducedinV}) we use the
recurrence relations (\ref{line1v4}),(\ref{line2v4}), 
(\ref{line3alter}), (\ref{v4line4}). After some additional algebra a 
relatively cumbersome formula is obtained:
\bea
&&(3d-10)(d-3)^2 m_4^2 V_{1111}^{(d)}=\frac{1}{4q^2}\left\{
 (3d-10)\Delta_{134}[\Delta_{246}
+u_{246}({\bf 2^-}-{\bf 4^-})]
\right.
\non \\
&& \non \\
&&~~~\left.
-8 (d-4)^2 q^2 m_3^2 m_4^2{\bf 4^-}
\right\}V_{1111}^{(d-2)}
\non \\
&& \non \\
&&~~~+\left\{ (d-3)m_4^2
[4m_1^2 (d-4){\bf 1^+}{\bf 1^+}
-8(d-3)m_2^2{\bf 2^+}{\bf 2^+}+(3d-10)(d-3) {\bf 2^+}]
                    \right.
\non \\
&& \non \\
&&~~~+u_{624}[\frac12(3d-10)u_{134}
( {\bf 1^+}\!- {\bf 3^+})\!+(d-2)(d-3)m_4^2 {\bf 1^+}]
 {\bf 2^+}
 \non \\
 && \non \\
&&~~~ \left.
 -m_4^2[(d-4)^2u_{134} {\bf 2^+}
 +(3d-10)u_{413} {\bf 3^+}] {\bf 1^+}
 \right\} {\bf 4^-} V_{1111}^{(d)}
\non \\
&& \non \\
&&~~~-(d-3)m_4^2\left\{4(d-4)m_1^2{\bf 1^+}
+(d-2) [u_{624}{\bf 2^+}-2m_4^2{\bf 4^+}]
       \right\}{\bf 1^+}{\bf 3^-}V_{1111}^{(d)}
\non \\
&& \non \\
&&~~~+\frac12(3d-10)u_{134}[u_{624}
{\bf 2^+}\!-2m_4^2{\bf 4^+}]({\bf 1^-}{\bf 3^+}
\!-{\bf 1^+}{\bf 3^-})V_{1111}^{(d)}.
\label{vddm2}
\eea
Notice that all terms with $V^{(d)}_{1111}$ on the right-hand side
of (\ref{vddm2}) are accompanied by shift operators contracting
one of the lines i.e. producing simpler integrals. We kept 
$V^{(d)}_{1111}$ for compactness of the formula, instead of 
specifying particular integrals obtained from it by contracting
lines. It is worthwhile to note that in the relation (\ref{vddm2})
when $\Delta_{134}=0$ or $\Delta_{246}=0$, the term 
$V^{(d-2)}_{1111}$ with uncontracted lines drops out and we obtain
expression for $V^{(d)}_{1111}$  in terms of simpler integrals.

Recall that if $m_4^2=0$, all $V^{(d)}$ for arbitrary $d$ are
reducible to integrals $J^{(d)}$ plus simpler ones and therefore in
this case the problem of reducing $d$ is transferred to simpler 
integrals.

Concluding this subsection it is worthwhile to mention that in the 
general mass case, after applying appropriate recurrence relations,
the only $V^{(d)}$ integral which will remain is $V^{(d)}_{1111}$.
This will be the second integral in our minimal basis of integrals.
In fact in calculating diagrams with the topology given in Fig.1.
four integrals $V^{(d)}_{1111}$ corresponding to different mass
distributions will occur (see Appendix ). All these integrals are
to be included in the basis.
\subsection{Integrals with three propagators $J^{(d)}$}

The remaining nontrivial integrals $J^{(d)}$ (see Fig.3), which are 
of interest in this subsection, are rather symmetrical though the 
corresponding recurrence relations are more cumbersome than in the
previous cases.

\begin{center}
\begin{picture}(140,100)(0,0)
\CArc(70,50)(25,0,180)
\CArc(70,50)(25,180,360)
\Line(30,50)(110,50)
\Vertex(45,50){2}
\Vertex(95,50){2}
\Text(45,72)[]{$1$}
\Text(70,58)[]{$2$}
\Text(70,35)[]{$3$}
\Text(70,0)[]{${\rm Fig.~3.~Diagrams~ with~three~lines}$}
\end{picture}
\end{center}
\vspace{3mm}
To reduce integrals with three lines to a basic set of integrals two
kind of recurrence relations are needed. The first one reduces the 
sum of indices of the integral if at least two lines have indices
exceeding 1. The relation applicable when $\nu_{1} > 1$ and
$\nu_{2} > 1$ has the form:
\bea
&&2 \nu_1\nu_2 D_{123} {\bf1^+}{\bf2^+}J^{(d)}_{\nu_1 \nu_2 \nu_3}
(q^2)=\left[ 2\nu_1h_{123}{\bf 1^+}+2\nu_2h_{213}{\bf 2^+}
 +4\nu_3 m_3^2 \sigma_{123} {\bf 3^+}
\right.\nonumber \\
&&\nonumber \\
&&~~~~~~
 +\nu_2 \nu_3 m_3^2\phi_{213}{\bf1^-}{\bf2^+}{\bf3^+}
 +\nu_1 \nu_3 m_3^2\phi_{123}{\bf1^+}{\bf2^-}{\bf3^+}
 -2\nu_1\nu_2 \rho_{123} {\bf1^+}{\bf2^+}{\bf3^-}  \nonumber \\
&&\nonumber \\
&&~~~~~~~~\left.
+\frac12\Sigma_3(d-\nu_1-\nu_2-\nu_3)\phi_{321}
 \right]J^{(d)}_{\nu_1 \nu_2 \nu_3}(q^2),
\label{a1a2}
\eea
where
\bea
&&D_{ijk}=q^8-4q^6(m_i^2+m_j^2+m_k^2)+q^4[6( m_i^4+m_j^4+m_k^4) \non \\
&&~~~~ +4(m_i^2 m_j^2+ m_k^2 m_i^2+m_k^2 m_j^2)]
-4q^2[m_i^6+m_j^6+m_k^6-m_i^2( m_j^4+m_k^4)
\nonumber \\
&&~~~~
 -m_j^2(m_i^4+m_k^4)
 -m_k^2( m_i^4+m_j^4)+10 m_i^2 m_j^2 m_k^2]+\Delta^2_{ijk},
\\
\nonumber \\
&&\rho_{ijk}=-\frac14 \frac{\partial D_{ijk}}{\partial q^2}=
-q^6+3q^4(m_i^2+m_j^2+m_k^2)
\nonumber \\
&&~~~~-q^2[3(m_i^4+m_j^4+m_k^4)
+2(m_i^2m_j^2+m_i^2m_k^2+m_j^2m_k^2)]
+m_i^6+m_j^6+m_k^6
\nonumber \\
&&~~~~
\!-\!m_i^2(m_j^4+m_k^4)\!-\!m_j^2(m_i^4
+m_k^4)\!-\!m_k^2(m_i^4+m_j^4)\!+\!10m_i^2m_j^2m_k^2,
\\
&&\nonumber \\
&&\phi_{ijk}=\frac12 \frac{\partial}{\partial m_i^2}
 \left( \frac{\partial}{\partial m_i^2}
       +\frac{\partial}{\partial m_j^2}
       +\frac{\partial}{\partial m_k^2}
       +\frac{\partial}{\partial q^2} \right) D_{ijk}=
\nonumber \\
&&~~~~4[q^4+2q^2(m_i^2-m_j^2-m_k^2)+(m_j^2-m_k^2)^2
+m_i^2(2m_j^2+2m_k^2-3m_i^2)]\non \\
&&~~~~=4[q^4+2q^2u_{ijk}+u_{ijk}u_{jik}+
u_{ijk}u_{kij}-u_{jik}u_{kij}],\\
&& \nonumber \\
&&\sigma_{ijk}=-\frac14(d-\nu_i-2\nu_j)\phi_{ijk}
               -\frac14(d-2\nu_i-\nu_j)\phi_{jik} \nonumber \\
&&~~~~~~~~~-\frac14(2d-2\nu_i-2\nu_j-\nu_k-1)\phi_{kij}, \\
&&\nonumber \\
&&h_{ijk}=-\frac12(d-2\nu_j-\nu_k)m_k^2 \phi_{ijk}
          -\frac12(2d-\nu_i-2\nu_j-2\nu_k-1)m_i^2\phi_{kij}
	  \nonumber \\
&& \nonumber \\
&&~~~~~~~~~~+(d-\nu_j-2\nu_k)\rho_{ijk}.
\eea
Two more recurrence relations follow from (\ref{a1a2}) by
the interchanges
\bea
&&{\bf 1^+ 3^+}J^{(d)}:
~~\nu_2,m_2,{\bf 2^{\pm}} \leftrightarrow \nu_3,m_3,{\bf 3^{\pm}},
\nonumber \\
&&{\bf 2^+ 3^+}J^{(d)}:
~~\nu_1,m_1,{\bf 1^{\pm}} \leftrightarrow \nu_3,m_3,{\bf 3^{\pm}}.
\label{interJ}
\eea
The above recurrence relations are of no use if two lines have an 
index equal to one. In this case the second set of recurrence 
relation is needed. To reduce the index of the first line the 
following relation applies:
\bea
\label{a1a1}
2 m_1^2 D_{123}\nu_1(\nu_1+1) {\bf1^+}{\bf1^+}J^{(d)}_{\nu_1 \nu_2 \nu_3}
(q^2)=\left\{ -\Sigma_3 (d-\nu_1-\nu_2-\nu_3)
\rho_{123}~~~~~~~~~~~~~~~~~~~~
 \right.
\nonumber \\
\non \\
+m_2^2m_3^2\phi_{123}\nu_2 \nu_3 {\bf1^-}{\bf2^+}{\bf3^+}
 +m_1^2 m_3^2\phi_{213}\nu_1 \nu_3{\bf1^+}{\bf2^-}{\bf3^+}
 +m_1^2 m_2^2\phi_{312}\nu_1 \nu_2{\bf1^+}{\bf2^+}{\bf3^-}
\nonumber \\
\non \\
\left.+ (d-2\!-2\nu_1)D_{123}\nu_1 {\bf 1^+}\!+
m_1^2 S_{123}\nu_1{\bf 1^+}\!+m_2^2 S_{213}\nu_2{\bf 2^+}\!+
m_3^2 S_{312}\nu_3 {\bf 3^+}\!\right\}\!J^{(d)}_{\nu_1 \nu_2 \nu_3}
(q^2),
\eea
where
\bea
&&S_{ijk}=-(d-2\nu_j-\nu_k)m_k^2\phi_{jik}
          -(d-\nu_j-2\nu_k)m_j^2\phi_{kij}
\nonumber \\
&&~~~~~~~~~~+2(2d-\nu_i-2\nu_j-2\nu_k-1) \rho_{ijk}.
\eea
Recurrence relations for the reduction of $\nu_{2,3}$ can be easily
deduced from (\ref{a1a1}) by symmetrical interchanges similar to 
(\ref{interJ}).

We see from (\ref{a1a1}) that if two indices are equal to 1 then it
is impossible to reduce the third index to 1. This means that our
minimal set of integrals will include not only $J^{(d)}_{111}$ 
but also $J^{(d)}_{211}$, $J^{(d)}_{121}$ and $J^{(d)}_{112}$.
Notice that the last three integrals are just derivatives
of $J^{(d)}_{111}$ with respect to different masses. From Fig. 1 we
see that there are two possibilities to obtain the integrals
$J^{(d)}$. One possibility corresponds to a contraction of the lines
1 and 4 and the other one to a  contraction of the lines 2 and 3.
Altogether, the representation of the integrals $J^{(d)}(q^2)$ 
requires us to include eight integrals in our minimal basis (see
Appendix).

For arbitrary $q^2$  the second set of recurrence relations needs
a modification when  some masses are equal to zero. If, for instance,
$m_1^2=0$,  then substituting this value into Eq. ($\ref{a1a1}$) we
deduce the simpler relation:
\bea
&&\nu_1 (d-2\nu_1-2) \Delta_{236}
{\bf1^+}J^{(d)}_{\nu_1 \nu_2 \nu_3}(q^2)
=\left\{-4m_2^2m_3^2\nu_2\nu_3{\bf 1^-2^+3^+} \right.
\nonumber \\
&&\nonumber \\
&&~~~~~~~~
+2m_2^2[(q^2-m_2^2)(2d-2\nu_1-\nu_2-2\nu_3-1)
-m_3^2(2\nu_1-\nu_2-1)]\nu_2 {\bf 2^+}
\nonumber \\
&&\nonumber \\
&&~~~~~~~~
+2m_3^2[(q^2-m_3^2)(2d-2\nu_1-2\nu_2-\nu_3-1)
-m_2^2(2\nu_1-\nu_3-1)]\nu_3 {\bf 3^+}
\nonumber \\
&&\nonumber \\
&&~~~~~~~~\left.-\Sigma_3(d\!-\nu_1-\nu_2-\nu_3)u_{623}
 \right\}J^{(d)}_{\nu_1 \nu_2 \nu_3}(q^2),
\eea
which can be used to lower the exponent of the massless propagator
down to 1. Therefore, in this case, the number of basic integrals
will be lower than in the  general case. Each massless line reduces
the number of basic $J^{(d)}$ type integrals by 1.

Turning next to $d$- recurrence relations, we note that the reduction
of $d$ for $J^{(d)}$ is a bit simpler than that of $V^{(d)}$.
The main reason is the existence of systematic recurrence relations
which reduce simultaneously some indices and $d$. If two indices,
say $\nu_1>1$ and $\nu_2 >1$, one can use the relation:
\bea
&&q^2(\Sigma_3+2)\nu_1\nu_2{\bf 1^+}{\bf 2^+}
J^{(d+2)}_{\nu_1 \nu_2 \nu_3}(q^2)
=\left\{q^2(d-2\nu_3)+m_1^2(d-\nu_1-2\nu_2)
 \right.
\non \\
&& \non \\
&&~~~~+m_2^2(d-2\nu_1-\nu_2)-2m_3^2(d-\nu_1-\nu_2-\nu_3)
-2m_3^2(q^2-m_3^2)\nu_3{\bf 3^+}
\non \\
&& \non \\
&&~~~~+m_1^2(q^2-m_1^2-3m_2^2+3m_3^2
-{\bf 2^-}+{\bf 3^-})\nu_1{\bf 1^+} \non \\
&& \non \\
&&~~~~\left.+m_2^2(q^2-3m_1^2-m_2^2+3m_3^2
-{\bf 1^-}+{\bf 3^-})\nu_2{\bf 2^+}
  \right\}J^{(d)}_{\nu_1 \nu_2 \nu_3}(q^2).
\label{dreda1a2}
\eea
Relations for ${\bf 1^+ 3^+}J^{(d+2)}$, ${\bf 2^+ 3^+}J^{(d+2)}$ 
are derivable from (\ref{dreda1a2}) by performing appropriate
interchanges (\ref{interJ}).
Relations reducing the index of one line look more complicated
\bea
&&-q^2m_1^2(\Sigma_3+2)\nu_1(\nu_1+1){\bf 1^+}{\bf 1^+}
J^{(d+2)}_{\nu_1 \nu_2 \nu_3}(q^2)
+\frac12 q^2(\Sigma_3+2)(\Sigma_3+4)\nu_1{\bf 1^+}
J^{(d+2)}_{\nu_1 \nu_2 \nu_3}(q^2)= \non \\
&&~~~~\left\{
 -\frac12q^4(d+2\nu_1-2\nu_2-2\nu_3)
+\frac12q^2[m_1^2(6\nu_1+6\nu_2+6\nu_3-7d)
       \right. \non \\
&& \non \\
&&~~~~+m_2^2(7d-2\nu_1-4\nu_2-10\nu_3)
      +m_3^2(7d-2\nu_1-10\nu_2-4\nu_3)]
\non \\
&& \non \\
&&~~~~+m_2^4(d-2\nu_1-\nu_2)+m_3^4(d-2\nu_1-\nu_3)
+m_1^2m_2^2(d-\nu_1-2\nu_2)
\non \\
&& \non \\
&&~~~~+m_1^2m_3^2(d-\nu_1-2\nu_3)-4m_2^2m_3^2(d-\nu_1-\nu_2-\nu_3)
\non \\
&& \non \\
&&~~~~-m_1^2[2q^4-q^2(2m_1^2+m_2^2+m_3^2)
 +m_1^2(m_2^2+m_3^2)+3(m_2^2-m_3^2)^2] \nu_1 {\bf 1^+}
\non \\
&& \non \\
&&~~~~+m_2^2[q^4+(q^2-m_2^2)(3m_1^2-5m_3^2+{\bf 1^-}-{\bf 3^-})
 -m_2^4]\nu_2 {\bf 2^+}
\non \\
&& \non \\
&&~~~~+m_3^2[q^4+(q^2-m_3^2)(3m_1^2-5m_2^2+{\bf 1^-}-{\bf 2^-})
 -m_3^4]\nu_3 {\bf 3^+}
\non \\
&& \non \\
&&~~~\left.
-\frac12q^2(\Sigma_3+2)({\bf 1^-}-{\bf 2^-}-{\bf 3^-})
-m_1^2(m_2^2-m_3^2)\nu_1{\bf 1^+}({\bf 2^-}-{\bf 3^-})
\right\}J^{(d)}_{\nu_1 \nu_2 \nu_3}(q^2).
\label{dreda1a1}
\eea
Two additional relations follow from (\ref{dreda1a1}) by interchanges
similar to (\ref{interJ}).

If $m_1=0$  Eq.(\ref{dreda1a1}) takes the simpler form
\bea
&&
q^2(\Sigma_3+2)(\Sigma_3+4)\nu_1{\bf 1^+}
J^{(d+2)}_{\nu_1 \nu_2 \nu_3}(q^2)=
\left\{
 -q^4(d+2\nu_1-2\nu_2-2\nu_3)
\right.\non \\
&&  \non \\
&&~~~~+q^2[m_2^2(7d-2\nu_1-4\nu_2-10\nu_3)
      +m_3^2(7d-2\nu_1-10\nu_2-4\nu_3)]  \non \\
&& \non \\
&&~~~~+2m_2^4(d-2\nu_1-\nu_2)+2m_3^4(d-2\nu_1-\nu_3)
 -8m_2^2m_3^2(d-\nu_1-\nu_2-\nu_3)\non \\
&& \non \\
&&~~~~+2m_2^2[q^4-(q^2-m_2^2)(5m_3^2-{\bf 1^-}+{\bf 3^-})
 -m_2^4]\nu_2 {\bf 2^+}\non \\
&& \non \\
&&~~~~+2m_3^2[q^4-(q^2-m_3^2)(5m_2^2-{\bf 1^-}+{\bf 2^-})
 -m_3^4]\nu_3 {\bf 3^+}\non \\
&& \non \\
&&~~~\left.-q^2(\Sigma_3+2)({\bf 1^-}-{\bf 2^-}-{\bf 3^-})
\right\}J^{(d)}_{\nu_1 \nu_2 \nu_3}(q^2).
\eea

Again it is  always possible to reduce $J^{(d)}$ with different
shifts in $d$ to a basic sets of integrals in different dimensions
and then to transform all master integrals to the generic dimension.
The relation connecting $d-2$ and $d$ dimensional integrals
$J_{\nu_1 \nu_2 \nu_3}^{(d)}(q^2)$ follow  from the differential
relationship given in Ref.\cite{connection}. Here we give it in
terms of shift operators:
\be
J_{\nu_1 \nu_2 \nu_3}^{(d-2)}(q^2)=(
\nu_1 \nu_2 {\bf 1^+}{\bf 2^+}
+\nu_1\nu_3{\bf 1^+}{\bf 3^+}
+\nu_2\nu_3{\bf 2^+}{\bf 3^+})J_{\nu_1 \nu_2 \nu_3}^{(d)}(q^2).
\label{Jconnection}
\ee
The right-hand side of this equation can be simplified by using
the recurrence relations  given previously in this subsection. 
As we already mentioned $J^{(d)}_{111}$ and
$\partial_iJ^{(d)}_{111}$ are elements of our minimal set of basic
integrals. For practical applications, however, it  useful to use
separate formulas for the reduction of $J^{(d)}_{111}$ and
$\partial_iJ^{(d)}_{111}$. Substituting $\nu_i=1$ into
Eq.(\ref{Jconnection}) and simplifying
the right-hand side we obtain the relation:
\bea
&&D_{123} J_{111}^{(d-2)}(q^2)=\non \\
&&~~~(d-3)\left \{
(3d-8)[3q^4-2q^2 t_1-\Delta_{123}]
+\tau_1 {\bf 1^+}+\tau_2{\bf 2^+}+\tau_3{\bf 3^+}
\right\} J_{111}^{(d)}(q^2) \non \\
&&~~~+\frac14
[(\partial_1 D_{123})J^{(d)}_{022}(q^2)
+(\partial_2 D_{123})J^{(d)}_{202}(q^2)
+(\partial_3 D_{123})J^{(d)}_{220}(q^2)],
\label{jdminus2}
\eea

where
\bea
&&D_{123}=
(q^2-r_1)(q^2-r_2)(q^2-r_3)(q^2-r_4)=q^8-4t_1q^6+2t_2q^4-4t_3q^2+t_4,
\non \\
&&\tau_i=-2(q^2-m_i^2)[q^4-2q^2(t_1-4m_i^2)+\Delta_{123}] \non \\
&&r_1=(m_1+m_2+m_3)^2,~~~~r_2=(m_1+m_2-m_3)^2, \non \\
&&r_3=(m_1-m_2+m_3)^2,~~~~r_4=(m_1-m_2-m_3)^2, \non \\
&&t_1=m_1^2+m_2^2+m_3^2= -u_{123}-u_{213}-u_{312}, \non \\
&&t_2=3(m_1^4+m_2^4+m_3^4)+2(m_1^2m_2^2+m_1^2m_3^2+m_2^2m_3^2)\non \\
&&~~~~=3(u_{123}u_{213}+u_{123}u_{312}+u_{213}u_{312})
      +2(u_{123}^2+u_{213}^2+u_{312}^2),
\nonumber \\
&&t_3=m_1^2(m_1^4-m_2^4-m_3^4)+m_2^2(m_2^4-m_1^4-m_3^4)
 +m_3^2(m_3^4-m_1^4-m_2^4)\nonumber \\
&&~~~~~~~~~~~~~~~~~~~~~~~~~~~~~~~~~~~~~~~ +10m_1^2m_2^2m_3^2
\non \\
&&~~~~=-u_{123}u_{213}u_{312}-\!u_{123}^2u_{213}
                             -\!u_{123}^2u_{312}
			     -\!u_{213}^2u_{123}
			     -\!u_{213}^2u_{312}
			     -\!u_{312}^2u_{123}
			     -\!u_{312}^2u_{213},\non \\
&& \non \\
&&t_4=\Delta_{123}^2.
\eea
By differentiating ($\ref{jdminus2}$) with respect to masses,
and upon simplifying the right-hand side, we  get a system of four
equations connecting $J^{(d-2)}_{111}$, $\partial_i J^{(d-2)}_{111}$
with $J^{(d)}_{111}$, $\partial_i J^{(d)}_{111}$. The solution of
this system is straightforward but tedious and has the form
\bea
&&3q^2(d-3)(d-4)(3d-8)(3d-10)J^{(d)}_{111}(q^2)=
\left\{ (d-4)^2q^6 \right. \non \\
&&~~~-2q^4 t_1(d-4)(6d-23)+q^2\left[ 5t_1^2(15d^2-117 d+224)
 -t_2(42d^2-331d+640)\right] \non \\
&&~~~-\frac14(d-5)\left[t_3 \left(27d-90\right)
-t_1t_2\left(3d-2 \right)
-2t_1^3\left(5d-26\right) \right]
 \non \\
&&~~~\left.+f(m_1^2,m_2^2,m_3^2){\bf 1^+}
 +f(m_2^2,m_1^2,m_3^2) {\bf 2^+}
 +f(m_3^2,m_2^2,m_1^2) {\bf 3^+} \right\}
 J^{(d-2)}_{111}(q^2)
\non \\
&&~~~\non \\
&&~~+g(m_1^2,m_2^2,m_3^2)J^{(d-2)}_{220}(q^2)
     +g(m_1^2,m_3^2,m_2^2)J^{(d-2)}_{202}(q^2) \non \\
&&~~~\non \\
&&~~+g(m_2^2,m_3^2,m_1^2)J^{(d-2)}_{022}(q^2),
\eea
where
\bea
&&f(m_1^2,m_2^2,m_3^2)=m_1^2(q^2-m_1^2)\left\{ -2(d-4)q^4+q^2
\left[4t_1(5d-18)-24m_1^2(2d-7)\right] \right. \non \\
&&~~~\left.-2\left(4d-13\right)t_1^2+2(9d-31)t_2
-24m_2^2m_3^2(4d-13)-24m_1^4(2d-7) \right\},
\\
&&~~~\non \\
&&g(m_1^2,m_2^2,m_3^2)= \frac{m_1^2m_2^2}{(d-4)}
\left[4(d-4)q^4-4(7d-24)q^2(3m_3^2-2t_1) \right.\non \\
&&~~~\left.-t_1^2\left(23d-80\right)
+t_2\left(9d-32\right)-12m_3^4(d-4)+12m_1^2m_2^2(7d-24)
\right],
\eea

\bea
&&3q^2(d-3)(d-4)(3d-10)
{\bf 1^+}J_{111}^{(d)}(q^2)=
\non \\
&&~~~~\left\{-m_1^2[q^4(7d-24)
-2q^2((4d-15)t_1-(d-5)m_1^2)\right.
\non \\
&&~~~~+\frac32(d-5)t_1^2+\frac52(d-3)t_2-2(5d-17)m_1^4-2(13d-45)
m_2^2m_3^2 ] {\bf 1^+} \non \\
&&~~~~+2m_2^2(q^2-m_2^2)[(d-3)(q^2+m_2^2-5m_3^2)+(7d-25)m_1^2]
{\bf 2^+}\non \\
&& \non \\
&&~~~~+2m_3^2(q^2-m_3^2)[(d-3)(q^2-5m_2^2+m_3^2)+(7d-25)m_1^2]
{\bf 3^+}\non \\
&& \non \\
&&~~~-(d-3)(d-4)q^4+q^2[(7d-30)(d-3)t_1-(7d-31)(3d-10)m_1^2]
\non \\
&&\left.~~+\frac14(d-5)[(17d-66)t_1^2-(3d-14)t_2-4(3d-10)
(m_1^4+5m_2^2m_3^2)]\right\}J_{111}^{(d-2)}(q^2)
\non \\
&& \non \\
&&~~~~+\frac{2}{d-4}[(q^2+m_2^2)(7d-24)+(d-4)m_1^2-(5d-18)m_3^2]
m_1^2m_2^2J^{(d-2)}_{220}(q^2)\non \\
&&~~~~+\frac{2}{d-4}[(q^2+m_3^2)(7d-24)+(d-4)m_1^2-(5d-18)m_2^2]
m_1^2m_3^2J^{(d-2)}_{202}(q^2)\non \\
&&~~~~-\frac{4}{d-4}[(q^2+m_2^2+m_3^2)(d-3)+(d-4)m_1^2]
m_2^2m_3^2J^{(d-2)}_{022}(q^2).
\label{dJred}
\eea
Relations for ${\bf 2^+}J^{(d)}$ and ${\bf 3^+}J^{(d)}$ in terms
of $d-2$ dimensional integrals follow
from (\ref{dJred}) by substitutions similar to (\ref{interJ}).

\subsection{Two-loop bubble integrals}
Rather frequently we will encounter two-loop bubble integrals.
They can be treated as $V^{(d)}_{\nu_1 0 \nu_2 \nu_3}$ or as
a value of $J^{(d)}_{\nu_1 \nu_2 \nu_3}(q^2)$ at $q^2=0$.
 It is more reasonable to consider these integrals
separately. We found  several recurrence relations most
useful for practical applications:
\bea
&&(d-2)\nu_1{\bf 1^+} J_{\nu_1 \nu_2 \nu_3}^{(d)}(0)
=\left\{-u_{123}  -{\bf 1^-}+{\bf 2^-}+{\bf 3^-}
 \right\}J_{\nu_1 \nu_2 \nu_3}^{(d-2)}(0),
\non \\
&& \non \\
&&(d-2) \nu_2 \nu_3 {\bf 2^+}{\bf 3^+}J_{\nu_1 \nu_2 \nu_3}^{(d)}(0)
=\left\{-2m_1^2\nu_1{\bf 1^+}+(d-2-2\nu_1)
 \right\}J_{\nu_1 \nu_2 \nu_3}^{(d-2)}(0),
\non \\
&& \non \\
&&(d-2)(d-\nu_1-\nu_2-\nu_3)J_{\nu_1 \nu_2 \nu_3}^{(d)}(0)=
\non \\
&& \non \\
&&~~~~-\left\{ \Delta_{123}
      +u_{123}{\bf 1^-}
      +u_{213}{\bf 2^-}
      +u_{312}{\bf 3^-}
      \right\}J_{\nu_1 \nu_2 \nu_3}^{(d-2)}(0),
\non \\
&& \non \\
&&\Delta_{123}\nu_1{\bf 1^+}J_{\nu_1 \nu_2 \nu_3}^{(d)}(0)
 =\left\{u_{123}(d-\nu_1-2\nu_2)
  +2m_2^2(\nu_1-\nu_2) \right.
\non \\
&& \non \\
&&~~~~\left.
+u_{312} \nu_1{\bf 1^+}({\bf 2^-}-{\bf 3^-})
+2m_2^2  \nu_2{\bf 2^+}({\bf 1^-}-{\bf 3^-})\right\}
 J_{\nu_1 \nu_2 \nu_3}^{(d)}(0).
\label{equa91}
\eea
These relations along with those obtained from 
(\ref{equa91}) by making interchanges like (\ref{interJ})
allow one to reduce any $J^{(d+2l)}_{\nu_1 \nu_2 \nu_3}(0)$ with
integer $\nu_i \geq0$  and integer $l$ to a combination of 
$J^{(d)}_{111}(0)$ and  products of different one-loop tadpole
integrals. Calculating integrals (\ref{J})
we will encounter only two $J^{(d)}_{111}(0)$ with
different mass distributions (see Appendix).

In the case when $\Delta_{123}=0$ the recurrence relations are 
simpler. If all masses are different from zero, without loss of
generality, we can assume that $m_1^2=(m_2+m_3)^2$  and then by
using the relation
\bea
&&2m_2 m_3(m_2+m_3)(d-2\nu_1-2\nu_2-2\nu_3-1)
{\bf 1^+}J_{\nu_1 \nu_2 \nu_3}^{(d)}(0)=
\non \\
&&\non \\
&&~~\left\{m_2[(d-\nu_1-\nu_2-2\nu_3-1)({\bf 1^+}{\bf 2^-}-1)
+(\nu_1-\nu_2+1){\bf 1^+}{\bf 3^-}] \right.
\non \\
&&\non \\
&&~~\left.+m_3[(d\!-\nu_1\!-2\nu_2\!-\nu_3-1)
({\bf 1^+}{\bf 3^-}\!-1)+(\nu_1-\nu_3+1){\bf 1^+}{\bf 2^-}]
\right\}J_{\nu_1 \nu_2 \nu_3}^{(d)}(0).
\eea
the integral can be reduced to a combination of products of
one-loop bubble integrals.
If $m_1=0$ and $m_3=m_2$ then $J^{(d)}$ also can be reduced to
a product of one-loop tadpoles by applying the relation:
\bea
&&2m_2^2(d-2\nu_1-2)(d-2\nu_1-\nu_2-\nu_3-1)
(d-2\nu_1-\nu_2-\nu_3)
{\bf 1^+}J_{\nu_1 \nu_2 \nu_3}^{(d)}(0)=\non \\
&&\non \\
&&(d-2\nu_1-2\nu_3)(d-2\nu_1-2\nu_2)(d-\nu_1-\nu_2-\nu_3)
J_{\nu_1 \nu_2 \nu_3}^{(d)}(0).
\eea
This relation is in agreement with the exact result 
first obtained in Ref. \cite{Vladim}.

\subsection{One-loop integrals}
The reduction of one-loop integrals does not cause any serious
problems. By applying the recurrence relation
\bea
&&\Delta_{126}\nu_1 {\bf 1^+}G^{(d)}_{\nu_1 \nu_2}
=\nonumber \\
&&~~~~
 \left\{
   u_{612}[\nu_1 {\bf 1^+}{\bf 2^-} -d +  \nu_1+2\nu_2]
   +2m_2^2[\nu_2 {\bf 2^+}{\bf 1^-} -d + 2\nu_1+\nu_2]
 \right\}G^{(d)}_{\nu_1 \nu_2},
\label{nu1inG}
\eea
along with the relation for ${\bf 2^+}G^{(d)}_{\nu_1 \nu_2}$
which follows from (\ref{nu1inG}) by interchanging
$\nu_1, m_1, {\bf 1^{\pm}}$ and $\nu_2, m_2, {\bf 2^{\pm}}$
any $G^{(d)}_{\nu_1 \nu_2}(q^2)$ with integer $\nu_i >0$
can be reduced to a combination of $G^{(d)}_{11}(q^2)$
and tadpoles $G^{(d)}_{0\nu_2}(q^2)$ and $G^{(d)}_{\nu_1 0}(q^2)$.
Thus, one-loop integrals $G^{(d)}_{\nu_1 \nu_2}$ will require only
$G^{(d)}_{11}$ in our set of basic integrals. The reduction of scalar
integrals (\ref{J}) will produce two integrals $G^{(d)}_{11}$. 
One will be made from the lines 1 and 3 and the other one made from
the lines 2 and 4 (see Appendix). To lower the space-time dimension
of $G^{(d)}_{\nu_1 \nu_2}$ two relations can be used:
\bea
&&2q^2\nu_1{\bf 1^+}G^{(d)}_{\nu_1 \nu_2}
=\left\{u_{126}+{\bf 1^-}-{\bf 2^-}\right\}
G^{(d-2)}_{\nu_1 \nu_2}, \\
&& \non \\
&&2q^2(d-\nu_1-\nu_2-1)G^{(d)}_{\nu_1 \nu_2}
=\left\{\Delta_{126}+u_{126}{\bf 1^-}
            +u_{216}{\bf 2^-}\right\} G^{(d-2)}_{\nu_1 \nu_2}.
\eea

As in the case of the two-loop bubble integrals, it is convenient
to consider one-loop bubble integrals separately. In principle, one
can use the explicit expressions for one-loop tadpole integrals,
however, it is more efficient to treat them on equal footing with 
the other integrals i.e. to apply recurrence relations for their
evaluation. Only two recurrence relations are needed for all
kinds of reductions:
\bea
&&2m_1^2 \nu_1 {\bf 1^+}T_{\nu_1}^{(d)}(m_1^2)=
(d-2\nu_1)T_{\nu_1}^{(d)}(m_1^2) \\
&& \non \\
&&(d-2\nu_1)T_{\nu_1}^{(d)}(m_1^2)=-2m^2_1T_{\nu_1}^{(d-2)}(m_1^2)
\eea
Five integrals $T_1^{(d)}(m_i^2)$, $i=1,..,5$ have to be taken as
elements of our basic set of integrals. Different products of one
loop integrals $G^{(d)}_{11}$ and/or $T_1^{(d)}$ which are to be
included in the minimal set of basic integrals are given in the
Appendix.

\section{Summary and conclusions}

In the present article we  described a procedure for the systematic
reduction of two-loop diagrams to a set of basic ones. We have shown
how a scalar contribution of the form (\ref{J}), obtained from  any
two-loop propagator diagram, can be reduced to a sum over 30 basic
structures $I_j^{(d)}(q^2)$
\be
I(q^2)=\sum_{j=1}^{30} R_j(q^2,\{m_i^2\},d) I_j^{(d)}(q^2),
\ee
with  $R_j(q^2,\{m_i^2\},d)$ being ratios of polynomials in
$q^2, \{m_i^2\}$ and $d$. All $I_j^{(d)}(q^2)$ are pictured in the
Appendix. The choice of the elements of the basis is not unique. One
can take some other 30 independent functions. We hasten to remark
that in order to know the two-loop integrals of the basis one needs 
to deal with  only three integrals  $F^{(d)}_{11111}$,
$V^{(d)}_{1111}$ and $J^{(d)}_{111}(q^2)$. The other members will be
obtained from those by changing masses or by differentiating. 
Two-loop bubble integrals are just the value of $J^{(d)}_{111}(q^2)$
at $q^2=0$. The number of basic structures strongly depends on the
mass values. If some masses are equal to zero or there are equal 
masses then the number of basic structures substantially diminishes.
For example, in the case of QED the number of relevant two-loop basic
integrals for the photon propagator is 2, and  for the fermion
propagator it is 5.

The evaluation of integrals from the basis is a separate problem.
Two-loop bubble and one-loop integrals are in a sense trivial. 
Analytic expressions for them are widely known (see for example,
\cite{BoDa}, \cite{DT}, \cite{JJ}).  As for the nontrivial two-loop
integrals the situation is the following. Integrals with three
propagators are related to the Lauricella functions \cite{BBBS}.
As was shown in \cite{BBBBW} integral with four propagators
$V_{1111}^{(d)}$ can be written in terms of multiple hypergeometric
series. Similar  series  for $F_{11111}^{(d)}$ with arbitrary
masses is not yet known. In the general mass case a one-fold integral
 representation for $F_{11111}^{(d)}$ is given in \cite{BB}.
Master integrals for propagator integrals occurring in QED and QCD
were investigated in \cite{Br1}, \cite{BFT}. Some particular cases
of two-loop diagrams with massive  particles were considered in 
\cite{ST}. We can recommended the reader for the general mass
case to use either the one-fold integral  representation 
\cite{BB} or to use the method of Refs.\cite{FT}, \cite{Jochem}.

We implemented the recurrence relations presented in this article
in a FORM \cite{FORM} package.  In a general mass case, 
required execution time for different diagrams varies from 
several seconds to several minutes. The only difficulty we 
encountered in calculating diagrams with all masses different 
was the lengthy expressions we obtained for the coefficients 
$R_j(q^2,\{m_i^2 \},d)$ and which are difficult to simplify
by using FORM. We see two possibilities to solve this problem.
The simplest solution will be to use other computer algebra system,
like Maple, Reduce, etc. to simplify  $R_j(q^2,\{m_i^2 \},d)$.
The other solution would be to assign numerical values to the
masses and then to perform calculations using FORM. To test the
algorithm we have reproduced two-loop result \cite{BFT} for the
photon self-energy.
 
Several remarks concerning the possibility to extend the presented
algorithm to a more complicated set of Feynman diagrams. Our 
preliminary investigation shows that a recurrence algorithm similar
to the one described in the present article can be worked out also
for the two-loop vertex (three-point) functions. Unfortunately, for
the general mass case kinematical determinants which appear
repeatedly  in recurrence relations are much more cumbersome than
for the propagator case. The implementation of these relations on the 
presently available computers is very problematic. For some
simplified kinematical cases, when say, some momenta or masses are equal to 
zero, or several masses are equal, the recurrence relations are 
simpler and can be implemented on a computer.

For integrals with several external momenta and several masses it may
be reasonable to solve the recurrence relations numerically.
It was done before  in \cite{Jochem} using multiprecision package
\cite{Bailey} for the small momentum expansion. It was noticed that in
this approach one needs  huge arrays that frequently cause problems
with a computer memory. However, if we do not consider small momentum
expansion and work with unexpanded diagrams then encountered indices
in integrals will be not so large  and therefore we will not meet
problems just mentioned.
\section{\em Acknowledgments. }
I am grateful to J.~Fleischer and F.~Jegerlehner for useful
discussions and for carefully reading the manuscript. I am 
thankful to  O.~Veretin for useful discussions  and for checks of
some formulae. I wish also to thank  York-Peng Yao for the  
clarification of some statements of the paper \cite{GY}.

\newpage
\section{Appendix}
In this appendix we present the complete set of integrals
resulting from the application of recurrence relations
for diagrams with arbitrary masses.

\begin{picture}(420,550)(0,-60)

\CArc(190,450)(30,0,180)
\CArc(190,450)(30,180,360)
\Line(190,420)(190,480)
\Line(140,450)(160,450)
\Line(220,450)(240,450)
\Vertex(160,450){1.75}
\Vertex(190,420){1.75}
\Vertex(190,480){1.75}
\Vertex(220,450){1.75}
\Text(160,470)[]{1}
\Text(220,470)[]{2}
\Text(160,430)[]{3}
\Text(220,430)[]{4}
\Text(197,450)[]{5}
\CArc(40,380)(20,0,180)
\CArc(40,380)(20,180,360)
\CArc(20,360)(20,0,90)
\Line(10,380)(20,380)
\Line(60,380)(70,380)
\Vertex(20,380){1.75}
\Vertex(40,360){1.75}
\Vertex(60,380){1.75}
\Text(40,409)[]{2}
\Text(25,358)[]{3}
\Text(55,358)[]{4}
\Text(40,380)[]{5}
\CArc(140,380)(20,0,180)
\CArc(140,380)(20,180,360)
\CArc(160,360)(20,90,180)
\Line(110,380)(120,380)
\Line(160,380)(170,380)
\Vertex(120,380){1.75}
\Vertex(160,380){1.75}
\Vertex(140,360){1.75}
\Text(140,409)[]{1}
\Text(125,358)[]{3}
\Text(155,358)[]{4}
\Text(140,380)[]{5}

\CArc(240,380)(20,0,180)
\CArc(240,380)(20,180,360)
\CArc(220,400)(20,270,360)
\Line(210,380)(220,380)
\Line(260,380)(270,380)
\Vertex(220,380){1.75}
\Vertex(260,380){1.75}
\Vertex(240,400){1.75}

\Text(230,409)[]{1}
\Text(250,409)[]{2}
\Text(250,358)[]{4}
\Text(240,380)[]{5}

\CArc(340,380)(20,0,180)
\CArc(340,380)(20,180,360)
\CArc(360,400)(20,180,270)
\Line(310,380)(320,380)
\Line(360,380)(370,380)
\Vertex(320,380){1.75}
\Vertex(360,380){1.75}
\Vertex(340,400){1.75}

\Text(330,409)[]{1}
\Text(350,409)[]{2}
\Text(350,358)[]{3}
\Text(340,380)[]{5}
%

\CArc(40,310)(20,0,180)
\CArc(40,310)(20,180,360)
\Line(10,310)(70,310)
\Vertex(20,310){1.75}
\Vertex(60,310){1.75}
\Text(54,334)[]{2}
\Text(42,320)[]{5}
\Text(38,299)[]{3}

\CArc(140,310)(20,0,180)
\CArc(140,310)(20,180,360)
\Line(110,310)(170,310)
\Vertex(120,310){1.75}
\Vertex(160,310){1.75}
\Vertex(140,330){2.25}
\Text(154,334)[]{2}
\Text(142,320)[]{5}
\Text(138,299)[]{3}

\CArc(240,310)(20,0,180)
\CArc(240,310)(20,180,360)
\Line(210,310)(270,310)
\Vertex(220,310){1.75}
\Vertex(260,310){1.75}
\Vertex(240,310){2.25}
\Text(254,334)[]{2}
\Text(242,320)[]{5}
\Text(238,299)[]{3}

\CArc(340,310)(20,0,180)
\CArc(340,310)(20,180,360)
\Line(310,310)(370,310)
\Vertex(320,310){1.75}
\Vertex(360,310){1.75}
\Vertex(340,290){2.25}
\Text(354,334)[]{2}
\Text(342,320)[]{5}
\Text(338,299)[]{3}

\CArc(40,250)(20,0,180)
\CArc(40,250)(20,180,360)
\Line(10,250)(70,250)
\Vertex(20,250){1.75}
\Vertex(60,250){1.75}
\Text(54,274)[]{1}
\Text(42,260)[]{5}
\Text(38,239)[]{4}

\CArc(140,250)(20,0,180)
\CArc(140,250)(20,180,360)
\Line(110,250)(170,250)
\Vertex(120,250){1.75}
\Vertex(160,250){1.75}
\Vertex(140,270){2.25}
\Text(154,274)[]{1}
\Text(142,260)[]{5}
\Text(138,239)[]{4}

\CArc(240,250)(20,0,180)
\CArc(240,250)(20,180,360)
\Line(210,250)(270,250)
\Vertex(220,250){1.75}
\Vertex(260,250){1.75}
\Vertex(240,250){2.25}
\Text(254,274)[]{1}
\Text(242,260)[]{5}
\Text(238,239)[]{4}

\CArc(340,250)(20,0,180)
\CArc(340,250)(20,180,360)
\Line(310,250)(370,250)
\Vertex(320,250){1.75}
\Vertex(360,250){1.75}
\Vertex(340,230){2.25}
\Text(354,274)[]{1}
\Text(342,260)[]{5}
\Text(338,239)[]{4}


\CArc(140,200)(20,0,180)
\CArc(140,200)(20,180,360)
\Line(120,180)(160,180)
\Line(140,180)(140,220)
\Vertex(140,180){1.75}
\Vertex(140,220){1.75}
\Text(115,200)[]{1}
\Text(135,200)[]{5}
\Text(155,200)[]{2}
\CArc(240,200)(20,0,180)
\CArc(240,200)(20,180,360)
\Line(220,180)(260,180)
\Line(240,180)(240,220)
\Vertex(240,180){1.75}
\Vertex(240,220){1.75}
\Text(215,200)[]{3}
\Text(235,200)[]{5}
\Text(255,200)[]{4}


\CArc(25,140)(15,0,180)
\CArc(25,140)(15,180,360)
\CArc(55,140)(15,0,180)
\CArc(55,140)(15,180,360)
\Line(0,140)(10,140)
\Line(70,140)(80,140)
\Vertex(10,140){1.75}
\Vertex(70,140){1.75}
\Vertex(40,140){1.75}

\Text(25,162)[]{1}
\Text(55,162)[]{2}
\Text(25,133)[]{3}
\Text(55,133)[]{4}


\CArc(120,140)(15,0,180)
\CArc(120,140)(15,180,360)
\CArc(155,155)(15,0,180)
\CArc(155,155)(15,180,360)
\Line(95,140)(105,140)
\Line(135,140)(170,140)
\Vertex(105,140){1.75}
\Vertex(155,140){1.75}
\Vertex(135,140){1.75}
\Text(120,162)[]{1}
\Text(120,133)[]{3}
\Text(155,162)[]{2}

\CArc(225,140)(15,0,180)
\CArc(225,140)(15,180,360)
\CArc(260,155)(15,0,180)
\CArc(260,155)(15,180,360)
\Line(200,140)(210,140)
\Line(240,140)(275,140)
\Vertex(210,140){1.75}
\Vertex(240,140){1.75}
\Vertex(260,140){1.75}
\Text(225,162)[]{1}
\Text(225,133)[]{3}
\Text(260,162)[]{4}

\CArc(325,140)(15,0,180)
\CArc(325,140)(15,180,360)
\CArc(360,155)(15,0,180)
\CArc(360,155)(15,180,360)
\Line(300,140)(310,140)
\Line(340,140)(375,140)
\Vertex(310,140){1.75}
\Vertex(340,140){1.75}
\Vertex(360,140){1.75}
\Text(325,162)[]{1}
\Text(325,133)[]{3}
\Text(360,162)[]{5}




\CArc(120,80)(15,0,180)
\CArc(120,80)(15,180,360)
\CArc(155,95)(15,0,180)
\CArc(155,95)(15,180,360)
\Line(95,80)(105,80)
\Line(135,80)(170,80)
\Vertex(105,80){1.75}
\Vertex(135,80){1.75}
\Vertex(155,80){1.75}
\Text(120,102)[]{2}
\Text(120,73)[]{4}
\Text(155,102)[]{1}

\CArc(225,80)(15,0,180)
\CArc(225,80)(15,180,360)
\CArc(260,95)(15,0,180)
\CArc(260,95)(15,180,360)
\Line(200,80)(210,80)
\Line(240,80)(275,80)
\Vertex(210,80){1.75}
\Vertex(240,80){1.75}
\Vertex(260,80){1.75}
\Text(225,102)[]{2}
\Text(225,73)[]{4}
\Text(260,102)[]{3}

\CArc(325,80)(15,0,180)
\CArc(325,80)(15,180,360)
\CArc(360,95)(15,0,180)
\CArc(360,95)(15,180,360)
\Line(300,80)(310,80)
\Line(340,80)(375,80)
\Vertex(310,80){1.75}
\Vertex(340,80){1.75}
\Vertex(360,80){1.75}
\Text(325,102)[]{2}
\Text(325,73)[]{4}
\Text(360,102)[]{5}


\CArc(25,30)(15,0,180)
\CArc(25,30)(15,180,360)
\CArc(60,30)(15,0,180)
\CArc(60,30)(15,180,360)
\Line(10,15)(75,15)
\Vertex(25,15){1.75}
\Vertex(60,15){1.75}
\Text(25,38)[]{1}
\Text(60,38)[]{2}


\CArc(120,30)(15,0,180)
\CArc(120,30)(15,180,360)
\CArc(155,30)(15,0,180)
\CArc(155,30)(15,180,360)
\Line(105,15)(170,15)
\Vertex(120,15){1.75}
\Vertex(155,15){1.75}
\Text(120,38)[]{1}
\Text(155,38)[]{4}

\CArc(225,30)(15,0,180)
\CArc(225,30)(15,180,360)
\CArc(260,30)(15,0,180)
\CArc(260,30)(15,180,360)
\Line(210,15)(275,15)
\Vertex(225,15){1.75}
\Vertex(260,15){1.75}
\Text(225,38)[]{1}
\Text(260,38)[]{5}

\CArc(325,30)(15,0,180)
\CArc(325,30)(15,180,360)
\CArc(360,30)(15,0,180)
\CArc(360,30)(15,180,360)
\Line(310,15)(375,15)
\Vertex(325,15){1.75}
\Vertex(360,15){1.75}
\Text(325,38)[]{2}
\Text(360,38)[]{3}

\CArc(25,-20)(15,0,180)
\CArc(25,-20)(15,180,360)
\CArc(60,-20)(15,0,180)
\CArc(60,-20)(15,180,360)
\Line(10,-35)(75,-35)
\Vertex(25,-35){1.75}
\Vertex(60,-35){1.75}
\Text(25,-12)[]{2}
\Text(60,-12)[]{5}


\CArc(120,-20)(15,0,180)
\CArc(120,-20)(15,180,360)
\CArc(155,-20)(15,0,180)
\CArc(155,-20)(15,180,360)
\Line(105,-35)(170,-35)
\Vertex(120,-35){1.75}
\Vertex(155,-35){1.75}
\Text(120,-12)[]{3}
\Text(155,-12)[]{4}

\CArc(225,-20)(15,0,180)
\CArc(225,-20)(15,180,360)
\CArc(260,-20)(15,0,180)
\CArc(260,-20)(15,180,360)
\Line(210,-35)(275,-35)
\Vertex(225,-35){1.75}
\Vertex(260,-35){1.75}
\Text(225,-12)[]{3}
\Text(260,-12)[]{5}

\CArc(325,-20)(15,0,180)
\CArc(325,-20)(15,180,360)
\CArc(360,-20)(15,0,180)
\CArc(360,-20)(15,180,360)
\Line(310,-35)(375,-35)
\Vertex(325,-35){1.75}
\Vertex(360,-35){1.75}
\Text(325,-12)[]{4}
\Text(360,-12)[]{5}
\Text(200,-80)[]{${\rm Fig.1.~ A~ complete ~set~ of~ basic~
 integrals}$}
\end{picture}

\end{document}